\documentclass[
reprint,
showpacs,
nofootinbib,
superscriptaddress,
longbibliography,
amsmath,amssymb,prd]{revtex4-2}

\usepackage{graphicx}%
\usepackage{dcolumn}%
\usepackage{siunitx}

\usepackage{bm}%
\usepackage{bbold}
\usepackage{braket}
\usepackage{esdiff}%
\usepackage{hyperref}%
\usepackage[mathlines]{lineno}%
\usepackage{subfigure}
\usepackage[skins,theorems]{tcolorbox}
\usepackage{soul}
\usepackage{mathtools}

\newcommand{\dv}[2]{\frac{\mathrm{d}#1}{\mathrm{d}#2}}

\begin{document}
\title{
Frequency shifts due to relativistic effects and retardation in continuous variable quantum key distribution}

\author{Emanuel Schlake}
\email{emanuel.schlake@zarm.uni-bremen.de}
\affiliation{ZARM, University of Bremen, Am Fallturm 2, 28359 Bremen, Germany}
\affiliation{Gauss-Olbers
Space Technology
Transfer Center, University of Bremen, 28359 Bremen, Germany}

\author{Jan P. Hackstein}
\email{jan.hackstein@zarm.uni-bremen.de}
\affiliation{ZARM, University of Bremen, Am Fallturm 2, 28359 Bremen, Germany}

\author{Roy Barzel}
\affiliation{ZARM, University of Bremen, Am Fallturm 2, 28359 Bremen, Germany}
\affiliation{Gauss-Olbers
Space Technology
Transfer Center, University of Bremen, 28359 Bremen, Germany}

\author{Eva Hackmann}
\email{eva.hackmann@zarm.uni-bremen.de}
\affiliation{ZARM, University of Bremen, Am Fallturm 2, 28359 Bremen, Germany}
\affiliation{Gauss-Olbers
Space Technology
Transfer Center, University of Bremen, 28359 Bremen, Germany}

\author{Dennis R\"atzel}
\email{dennis.raetzel@zarm.uni-bremen.de}
\affiliation{ZARM, University of Bremen, Am Fallturm 2, 28359 Bremen, Germany}
\affiliation{Gauss-Olbers
Space Technology
Transfer Center, University of Bremen, 28359 Bremen, Germany}

\author{Claus L\"ammerzahl} 
\affiliation{ZARM, University of Bremen, Am Fallturm 2, 28359 Bremen, Germany}
\affiliation{Gauss-Olbers
Space Technology
Transfer Center, University of Bremen, 28359 Bremen, Germany}

\begin{abstract}
Space-based quantum communication naturally involves satellites and ground stations exchanging optical signals at high altitudes and large relative velocities.
Starting from general relativistic considerations, we systematically separate the frequency shift into longitudinal Doppler contributions, relativistic corrections, and corrections from the propagation delay (retardation).
We find the relativistic corrections to the Keplerian satellite orbits to be negligible on the considered timescale, compared to the gravitational and special relativistic time dilation contributions to the frequency shift.
Somewhat surprisingly, we find the contribution from the retardation effect to be on the same order of magnitude as the relativistic contributions.
To analyze the significance of these effects, we investigate secret key rates for a continuous variable quantum key distribution protocol for various configurations of satellite orbits and ground stations.
We find that the corrections from relativistic effects and retardation significantly impact the communication performance and should be taken into account.
\end{abstract}

\maketitle

\clearpage
\newpage

\section{\label{sec:introduction} Introduction}

Quantum key distribution (QKD) promises unconditionally secure exchange of confidential information between two or more parties in the sense that the security of the protocol does not depend on the computational effort to break the key. This enables mitigation of the risk to classical encryption posed by the development of quantum computing resources
and has led to the vision of a global quantum communication network, the Quantum Internet \cite{kimble2008quantum, wehner2018quantum,quantum_internet_cacciapuoti_2020_computing,awschalom_internet_2020, Rohde_Internet_2021, azumaQuantumRepeatersQuantumInternet2023}. 

The natural carriers for quantum
information over large distances are photons as they propagate with 
the speed of light and are sufficiently isolated from their environment even in media. Moreover, they are already widely used in classical communication networks based on optical fibers and many existing technological solutions can be taken over to QKD. Still, the propagation of photons in optical fibers is limited to several hundreds of kilometers due to attenuation. One strategy for extending QKD to larger distances is the application of quantum repeaters \cite{briegel1998quantum,
jiang2009quantum,
sangouard2011quantum,
azuma2015all,
vinay2017practical,
bhaskar2020experimental}, whose functionality relies on pre-distributed entanglement and quantum memories \cite{sangouard2011quantum,
munro2015inside,
azuma2015all,
zwerger2018long,
su2018efficient%
}.
Another approach is the application of optical free-space communication, including satellite-based links \cite{boone2015entanglement,Sidhu2021adv,Lu_cao_micius_space_review_2022}. The combination of both approaches provides a viable path towards global quantum communication motivating many studies in the last decades \cite{bedington2017progress,
bourgoin2013comprehensive,
bonato2009feasibility,
villoresi2008experimental,
yin2013experimental,
vallone2015experimental,
dequal2016experimental,
carrasco2016leo,
takenaka2017satellite,wallnofer2022sim,
gundogan2021topical,
calderaro2018towards,
meister2025simulation}. Such considerations are supported by recent experimental breakthroughs, ranging from the first satellite-to-ground QKD implementation \cite{liao2017satellite,yin2017satellite} to intercontinental quantum-secured data transfer \cite{liao2017space,liao2018satellite}. For practical implementations that allow for high performance and stable data flow, space-qualified and robust optical components have to be developed. A promising way of implementation is offered by \textit{continuous-variable QKD} (CV-QKD) \cite{grosshans2002continuous,
ralph2000security,
grosshans2003quantum,
diamanti2015distributing,
laudenbach2018continuous}%
which uses approved standard telecommunication technology for state preparation and detection.
Several experimental demonstrations of CV-QKD in terrestrial applications revealed the potential of this technology for long-distance quantum communication at high key rates \cite{liao2017satellite,
yin2017satellite,
ren2017ground,
yin2017satellite2,
takenaka2017satellite} and some preliminary experimental studies have been performed on signal transmission along free-space and satellite-to-ground links \cite{heim2014atmospheric,gunthner2017quantum}.
Most of the theoretical studies on satellite-based free-space (CV-) QKD links focus on atmospheric \cite{liorni2019satellite,
semenov2012homodyne,
ruppert2019fading,
wang2018atmospheric} and geometric effects \cite{vasylyev2012toward,dequal2021feasibility}, such as satellite pointing errors, beam wandering or beam divergence due to atmospheric turbulence, which are undoubtedly the strongest decoherence sources distorting the performance of satellite-to-ground QKD protocols.

Another important effect that can significantly limit the performance of CV-QKD are frequency shifts of the signal with respect to emitter and observer, such as the Doppler shift due to their relative motion.
The resulting deformation of the signal spectrum relative to the reference spectrum in homodyne detection at the receiver leads to a mode mismatch
which acts as an effective loss
\cite{schlakePulseShapeOptimization2024, lordi2023_Quantum_theory_temporally_mismatched_homodyne}.
This is relevant in \emph{local} local oscillator (LLO) setups, where the reference signals are generated locally at emitter and receiver \cite{Soh_selfreferenced_cvqkd_llo_2015, Bing_Locallocaloscillator_CVQKD_2015, Huang_locallocaloscillator_cvqkd_2015}.
Since the effect increases for narrow-band signals, it becomes especially relevant in coherent quantum communication, where strong phase stability often requires very narrow-band signals
\cite{Soh_selfreferenced_cvqkd_llo_2015,Bing_Locallocaloscillator_CVQKD_2015,Huang_locallocaloscillator_cvqkd_2015,Liao2025,marieSelfcoherentPhaseReference2017a}
and quantum memories are in use.
In space-based operations, quantum memories typically need to be long-lived and therefore store narrow-band signals
\cite{gundogan2021topical, gundogan_2021_space_borne_quantum_memories, beavan_2013_spectral_filter_quantum_memory}.
For such spectrally narrow signals, relativistic frequency shifts such as the transverse Doppler shift and the general relativistic gravitational redshift can have a measurable effect on quantum communication tasks as shown in \cite{
bruschi2014spacetime,bruschi2014quantum,
kohlrus2017quantum,kohlrus2019quantum,liu2020characterization,Liu2022_gravitational}. To the best of our knowledge, all previous studies are restricted to very few special cases, for example, light signal exchange exclusively in the radial direction of the Earth. Here, we provide a characterization of relativistic effects on a practical satellite-based QKD task under realistic conditions
by resolving the time dependence of the dynamic configurations.

In the present paper, we investigate how the performance of the standard Gaussian modulated coherent state (GMCS) CV-QKD protocol \cite{grosshans2003quantum} is affected if relativistic and retardation effects, both on signals and satellites, are not compensated for. To this end, we find general expressions for the various relevant effects and exemplify them for some common satellite and ground station configurations. A major component of this work involves the determination of the signal linking a sender and a receiver, all moving on a curved space-time background. This is the well-known \textit{ emitter--observer problem} (EOP)  \cite{semerak2015approximating,viergutz1993image} for which generally no analytic solution is known. We then investigate the impact on
the so-called PLOB-bound \cite{pirandola2017fundamental}, which is a fundamental upper bound on the secret key rate of any bosonic two-way quantum communication protocol. Although this work focuses on point-to-point links between satellites and ground stations, such pairwise connections form the building blocks of more complex satellite constellations.
Therefore our study provides insight into which links should be prioritized to maximize overall communication performance.

From general relativistic considerations, general formulae for the frequency shift are derived
up to orders including $\mathcal{O}(v^2/c^2)$.
It is shown that the leading
order relativistic effects\footnote{The leading order relativistic effects are of order $v^2/c^2$. } are constituted solely by transversal Doppler shift and gravitational frequency shift, while relativistic orbital perturbation and effects on light propagation are of higher order
and do not affect communication performance significantly over time scales of a few orbital periods.
Additionally, we find that the change in frequency shift due to the light's propagation time compared to instantaneous propagation (retardation), contributes at the same order as the first order relativistic effects.
Therefore, this retardation must be taken into account.
We then compare the frequency shifts for Newtonian dynamics with simulations from a fully relativistic numerical framework recently introduced in \cite{hackstein2025, hackstein2024}.%

The work is organized as follows: In Sec. \ref{sec:cvqkd} we introduce the CV-QKD protocol under consideration and discuss the effect of frequency shift induced mode mismatch on the quantum channel's capacity.
In Sec. \ref{sec:freq_shift} the frequency shift in satellite communication is investigated and decomposed into contributions from relativistic effects, retardation and relativistic corrections to Keplerian orbital dynamics.
In Sec. \ref{sec:examples} we present examples of ground station-satellite and satellite-satellite communication links and discuss the present frequency shifts and the effect on the channel capacity.
The higher order corrections are numerically investigated and discussed in Sec. \ref{sec:higher_order}.
We conclude in Sec. \ref{sec:conclusion}.

\section{CV-QKD and mode match}\label{sec:cvqkd}

In CV-QKD, secure key generation relies on accurately measuring the quadratures of optical signals using a local oscillator as the phase reference.
When the local oscillator is generated locally by each communicating party (\emph{local} local oscillator, LLO), the security loopholes associated with the transmitted local oscillator are avoided \cite{Soh_selfreferenced_cvqkd_llo_2015, Bing_Locallocaloscillator_CVQKD_2015, Huang_locallocaloscillator_cvqkd_2015}.
The LLO design, however, requires precise mode matching between the local oscillator and signal states.
Here, we briefly introduce the considered CV-QKD protocol and how mode match enters as an effective transmissivity of the quantum channel, directly affecting the channel capacity.

To demonstrate the effect of mode match explicitly, we consider the CV-QKD protocol with Gaussian-modulated coherent states (GMCS) \cite{marieSelfcoherentPhaseReference2017a}.
Using her laser source and a modulator, Alice generates a train of signal and reference pulses.
The signal states are weak coherent states, $\ket{\alpha}$, carrying the quantum information. Alice randomly modulates either $x$ or $p$ quadrature (according to her local reference laser) as drawn from a Gaussian distribution \cite{grosshans2002continuous, grosshans2003quantum}.
The reference pulse is of high intensity and carries only information about the phase used by Bob to correct the phase of his local reference.
Having received the pulses, Bob performs homodyne detection using his local oscillator and measures, randomly, either $x$ or $p$ quadrature.
After measurements are complete, Bob can correct potential phase errors using the phase reference pulses Alice provided. The protocol ends with the usual sifting, privacy amplification and finally key extraction. A detailed review discussing this protocol can be found in e.g. \cite{weedbrook2012gaussian}.
The achievable secret key rate depends on the parameters of the considered optical quantum channel as well as the protocol itself. Secret key rates for several protocols were calculated for example in \cite{GP, pirandola2017fundamental}.
In satellite communication, major error sources include thermal noise, pointing errors, diffraction, and atmospheric extinction; these were investigated in detail in \cite{pirandolaLimitsSecurityFreespace2021}.
However, since in this article we focus on the relativistic mode mismatch between local oscillator and signal, we will not account for these additional errors individually but we include a baseline transmissivity of $\eta_{0} = 0.4$ to account for generic loss in the channel.

\begin{figure}[t]\centering
  \includegraphics[width=0.75\linewidth]{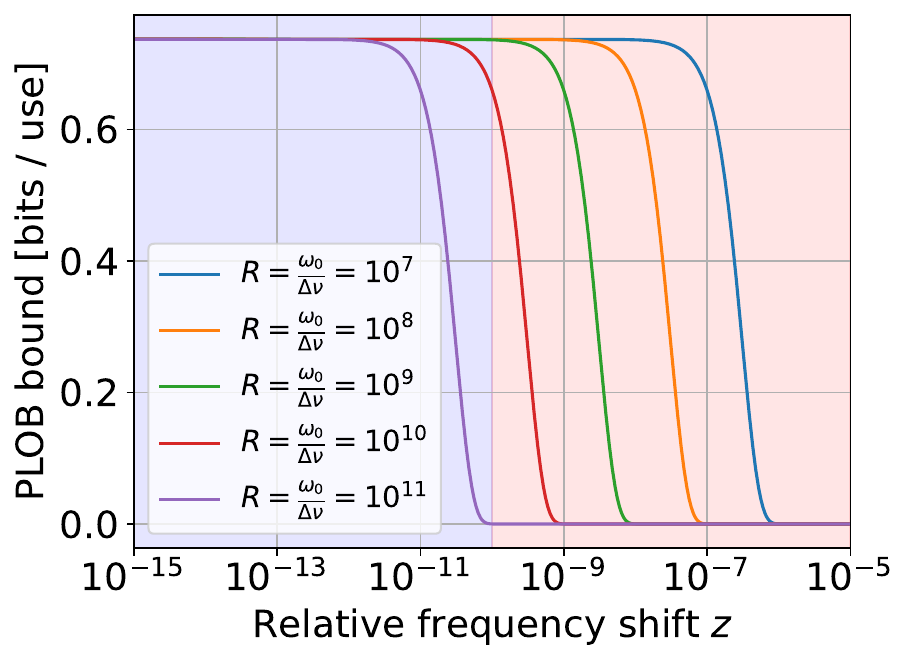}
  \caption{PLOB bound (Eq. \eqref{eq:cap_plob}) as  a function of the relative frequency shift for different values of peak-frequency-to-bandwidth ratio $R = \frac{\omega_{0}}{\Delta \nu}$.
  The channel transmissivity is given by the overlap of a Gaussian spectral amplitude with its frequency shifter counterpart according to Eq. \eqref{eq:wamb_gauss} as well as the assumed detector efficiency of $\eta_{0} = 0.4$.
  For relative frequency shifts smaller than the peak frequency to bandwidth ratio the PLOB bound is nearly constant at $ - \log_{2}(0.6) \approx 0.74$.
  At larger frequency shifts the PLOB bound drops exponentially and key rates become severely limited.
  Relative frequency shifts in the range $z \lesssim 10^{-10}$ are typical for (general) relativistic contributions for satellites in Earth's gravitational field while larger shifts are due to the longitudinal Doppler shift.
  }
\label{fig:PLOB}
\end{figure}

Signal and local oscillator are assumed to be in continuous mode coherent states (\cite{blowContinuumFieldsQuantum1990, Mandel_Wolf_1995, vanenkQuantumStatePropagating2001}) described by their normalized spectral amplitudes $F_{S}(\omega)$ and $F_{L}(\omega)$, respectively.
\begin{align}
    \label{eq:coherent_signal}
    \Ket{\alpha_{S}}_{S} &= \bigotimes_\omega \Ket{\alpha_{S} F_{S}(\omega)},
    \\
    \label{eq:coherent_LO}
    \Ket{\alpha_{L}}_{L} &= \bigotimes_\omega \Ket{\alpha_{L} F_{L}(\omega)},
\end{align}
where $\alpha_{S, L}$ is the complex amplitude with modulus $|\alpha_{S,L}|$ and phase $\theta_{S,L}$, respectively.
The number of photons in the coherent state has an expectation value given by the amplitude $\left< n_{S,L} \right> = | \alpha_{S, L}|^{2}$.
Given these states, we define the mode match (or overlap) between signal and local oscillator as
\begin{equation}
	\gamma := \int d\omega F_S(\omega) F^*_{L}(\omega).
\end{equation}
We will consider the Gaussian spectral amplitude shape throughout
\begin{equation}
    F(\omega)=\frac{1}{\sqrt[4]{2\pi\sigma^2}}e^{-\frac{(\omega-\omega_0)^2}{4\sigma^2}},
\end{equation}
where $\omega_0$ is the carrier frequency (peak frequency) and $\sigma$ is the width of the Gaussian, related to the bandwidth $\Delta \nu$ by $\Delta \nu = \sigma \sqrt{\ln 4}$.
The overlap between a Gaussian and its Doppler-shifted counterpart is given by \cite{bruschi2014spacetime, schlakePulseShapeOptimization2024}
\begin{equation}
    \label{eq:overlap_gauss}
    \gamma(z) = \sqrt{\frac{2(z+1)}{z (z+2)+2}} e^{ -\frac{\omega_0^2 z^2}{4 \sigma^2 (z (z+2)+2)}}.
\end{equation}%
For small Doppler shifts, $z \ll 1$, and large peak frequency to bandwidth ratio, $\omega_0 /\Delta \nu \gg 1$, this reduces to the following

\begin{equation}
    \label{eq:wamb_gauss}
    \gamma(z) = e^{-\frac{\omega_{0}^{2} z^2}{8 \sigma ^2}} = 2^{-\frac{R^{2}}{4} z^{2}}
    ,
\end{equation}
where $R = \omega_{0}/\Delta \nu$ is the peak-frequency-to-bandwidth ratio of the signal. The relation of $R$ to the PLOB bound is illustrated in Fig. \ref{fig:PLOB}.
This expression neglects the stretching of the wavepacket due to the Doppler shift \footnote{ For a discussion of the effect of the stretching on CV-QKD see \cite{schlakePulseShapeOptimization2024}.}.

At reception Bob performs balanced homodyne detection with a strong local oscillator of a random quadrature as the protocol dictates.
As was shown in \cite{schlakePulseShapeOptimization2024} the modulus of the overlap acts as an effective transmissivity $\eta_\gamma = | \gamma|$ and the overlap's argument as an additional phase $\Gamma = \arg \gamma$ of the associated quantum channel.
The effective channel is a composition of a lossy channel of transmissivity $\eta = \eta_0 \eta_\gamma$ and a phase shifter that changes the coherent states phase by $\Gamma$.
The induced phase error $\Gamma$, however, can be corrected after detection in post-processing \cite{marieSelfcoherentPhaseReference2017a}.
Since we are considering Gaussian spectral amplitudes for which the overlap is real, see Eq. \ref{eq:overlap_gauss}, there is no phase shift in this case.
Formally, the channel is then a pure loss channel given by
\begin{equation}
    \label{eq:loss_channel}
    \mathcal{L}_\eta: \Ket{\alpha} \mapsto \Ket{\eta \alpha}.
\end{equation}
The secret key and quantum capacity of the (otherwise ideal) lossy quantum channel with a loss of $1-\gamma$ is given by the PLOB bound \cite{pirandola2017fundamental}
\begin{equation}
	\label{eq:cap_plob}
	P(\mathcal{L}_{\eta}) = - \log_{2}(1-\eta), \quad \eta = \eta_{0} \eta_{\gamma}
\end{equation}

where $\eta_{\gamma} = | \gamma|$ is given by Eq. \eqref{eq:wamb_gauss}.
Having introduced the baseline transmissivity $\eta_0 = 0.4$, we avoid the divergence of the PLOB bound of the ideal channel ($\eta = 1$)
and the PLOB bound is instead limited by
\begin{equation}
	P(\eta = \eta_{0}) = -\log_{2}(0.6) \approx 0.74\,.
\end{equation}
For small frequency shifts $z\ll 1$, the PLOB bound can be approximated to be
\begin{equation}
	P(z) = -\log_{2} (1- \eta_{0}) - \frac{\eta_{0}}{1-\eta_{0}} \frac{R^2}{4} z^{2} + \mathcal{O}(z^{4})\,.
\end{equation}

\section{Frequency shift in satellite communication}
\label{sec:freq_shift}

In this section, we derive expressions for the relative frequency shift based on general relativistic considerations.
The emitter and receiver are assumed to move along timelike curves $\gamma_{A}(\tau_A)$ and $\gamma_{B}(\tau_B)$ with four-velocities $u_\mathrm{A}^{\mu}$ and $u_\mathrm{B}^{\mu}$, parametrized by their proper times $\tau_A$ and $\tau_B$, which coincide with the readings of clocks carried by the respective satellites.
Neglecting atmospheric effects (which will be relevant whenever ground stations are involved), 
light rays in vacuum follow null geodesics of the underlying spacetime and are
parametrized by some affine parameter \cite{MTW_gravitation,straumannGeneralRelativity2013}.
A geodesic's equation of motion is the geodesic equation
\begin{equation}
	\label{eq:geod}
	\ddot{x}^{\mu}+ \Gamma^{\mu}_{\alpha \beta} \dot{x}^{\alpha} \dot{x}^{\beta} = 0,
\end{equation}
where $\dot{x}^\mu$ denotes the derivative of the spacetime coordinate $x^{\mu}$ with respect to proper time $\dv{x^\mu}{\tau}$ or affine parameter $\dv{x^\mu}{s}$, respectively.
$\Gamma^{\mu}_{\alpha \beta}$ are the Christoffel symbols of the metric. We use the $(-,+,+,+)$ signature of the metric.
Throughout, Greek letters run over all spacetime indices, and we employ the Einstein summation convention.

\begin{figure}[t]\centering
  \includegraphics[width=0.8\linewidth]{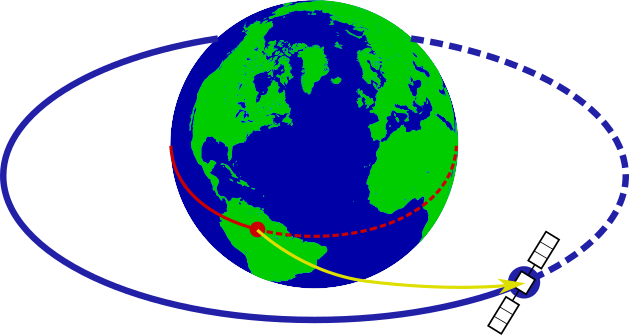}
  \caption{
	Sketch of the emitter--observer problem for a signal transmission from an equatorial ground station to an equatorial satellite.
  	The signal's propagation must account for the relative motion of the receiving satellite to ensure proper detection.
  	General relativistic effects on the light ray, such as gravitational bending and Shapiro delay, are present in principle but, as shown in the main text, are of lower order than the gravitational redshift and special relativistic Doppler shift.
}
\label{fig:EOPGeneral}
\end{figure}

The frequency of a light ray with
wave covector $k$, observed by some observer with four-velocity $u$, is given by the contraction of four-velocity and wave covector
$\omega = - k_\alpha u^\alpha$. %
The relative frequency shift $z$ between emitter, $u_{A}$, and receiver, $u_{B}$, is then in all generality \cite{kermack1934iv}
\begin{equation}
	\label{eq:freq_shift2}
	z = \frac{\omega_{A}}{\omega_{B}} - 1 = \frac{u_{A}^{\alpha} k_{\alpha} |_{A}}{u_{B}^{\beta} k_{\beta} |_{B}} - 1,
\end{equation} where $A$ and $B$ denote the events of emission and reception, respectively.
In a stationary spacetime the frequency shift can be factorized as \cite{philipp2023generalrelativisticchronometryclocks}
\begin{equation}
	z = \frac{\omega_{A}}{\omega_{B}} - 1 = \frac{u^{0}_{A}\vert_A}{u^{0}_{B}\vert_B} \left(\frac{1 + \frac{k_i v_A^i |_A}{k_0 c}}{1 + \frac{k_j v_{B}^j |_B}{ k_0 c} }\right) - 1,
\end{equation}
where $v^i$ are the spatial coordinate velocities, and Latin indices run over spatial components.
The first factor is the time dilation between emitter and receiver (i.e. special relativistic and gravitational time dilation) and the second factor is a longitudinal Doppler shift resulting from the relative coordinate velocities.
This, however, assumes that the light ray connecting emitting and receiving satellite has already been determined, which, even in flat spacetime, is a non-trivial problem: the so-called emitter--observer problem (EOP).
A sketch of the EOP for an uplink between a ground station and a satellite is shown in Fig. \ref{fig:EOPGeneral}.
The emitter--observer problem then consists of finding the light ray which connects a given emission event, $\gamma_{A}(\tau_\mathrm{em})$ with the timelike curve of the receiver.
This can be done either by solving for light rays on the basis of analytical approximations, as done e.g. in \cite{semerak2015approximating} for the Schwarzschild spacetime, or by integrating the geodesic equation numerically.
In this paper, we approach this problem in two ways: first, we split the contributions to the frequency shift into an instantaneous part, that involves only quantities evaluated at a fixed emitter time, and a retardation part, that estimates the impact of the relativistic propagation of the signal.
Second, 
the EOP was solved numerically for a number of test orbit setups by using the \textit{GREOPy} software package \cite{hackstein2025, hackstein2024};
for more information on the EOP and the numerical implementation,
see Appendix \ref{APP:EOPEA}.

In comparison to the Newtonian treatment, the relativistic treatment introduces a number of modifications to the satellite's and light's dynamics and frequency shifts.
Those include special relativistic effects such as corrections to the longitudinal Doppler shift, the transverse Doppler shift, gravitational redshift, light bending, Shapiro delay and corrections to Keplerian orbits.
In the following, we show that of these effects, only the corrections to Doppler shift, the gravitational redshift  and the retardation are the most relevant to space based communication scenarios.

\subsection{Relativistic contributions to the frequency shift}
\label{sec:z_rel_contr}

Here we derive contributions to the frequency shift as obtained from taking into account relativistic effects.
We assume that the light signal follows null geodesics of the underlying spacetime with tangent vector $k^{\alpha}$ as in geometrical optics \cite{MTW_gravitation}.
Given coordinates $x^{\alpha}$, the metric can then be written as a perturbation to the Minkowski metric $\eta_{\alpha \beta}$
\begin{equation}
	g_{\alpha \beta} = \eta_{\alpha \beta} + h_{\alpha \beta},
\end{equation} where we assume $h_{\alpha \beta}$ to be small.
The relativistic frequency shift for a signal exchange between two observers in a weak, stationary and axially symmetric field has been derived in \cite{linetTimeTransferFrequency2002}.
In the following, we shortly present parts of the results of \cite{linetTimeTransferFrequency2002} and make use of them in the remaining article.
Following \cite{linetTimeTransferFrequency2002}, 
the relative frequency shift up to order $\mathcal{O}(\frac{v^{2}}{c^{2}})$ is
\begin{equation}
	\label{eq:z_total}
    z = z_\mathrm{rel} + z_\mathrm{long},
\end{equation}
with
\begin{equation}
	\label{eq:z_long}
	\begin{aligned}
	z_\mathrm{long} =&%
	\frac{\bm{N_{AB}} \cdot (\bm{v_{B}} - \bm{v_{A}})}{c} 
		       \\&+ \frac{1}{c^{2}}\left( \bm{N_{AB}} \cdot (\bm{v_{B}} - \bm{v_{A}} )(\bm{N_{AB}} \cdot \bm{v_{B}}) \right)
    \end{aligned}
\end{equation}
and 
\begin{equation}
	\label{eq:z_rel}
    	z_\mathrm{rel} = \frac{1}{c^{2}} \bigg[
	\frac{1}{2}(v_{A}^{2} - v_{B}^{2}) 
	+ \left( \frac{GM}{r_{A}} -  \frac{GM}{r_{B}} \right)
	\bigg],
\end{equation}
where boldface letters denote three-vectors, $\bm{N_{AB}} = \frac{\bm{r_B} - \bm{r_A} }{|\bm{r_B} - \bm{r_A}|}$ is the normalized cartesian spatial coordinate difference between the emission and reception event
\footnote{Note that $N_{AB}$ is purely a coordinate difference vector
and does not include effects of light deflection.},
and $\bm{r_A}$, $\bm{r_B}$ denote the respective position vectors, while $r_A$, $r_B$ denote the corresponding radial coordinates.
Note that emission and reception event naturally occur at different coordinate times due to propagation delay.
In Earth-based satellite communication, the longitudinal Doppler shift typically reaches values of order $z_{\mathrm{long}} \sim 10^{-5}$ \cite{ali1998doppler}, whereas the relativistic contributions are of order $z_{\mathrm{rel}} \sim 10^{-10}$ \cite{Soffel_2003, ashby2003relativity}.

A discussion and numerical investigation of the higher order general relativistic effects such as the aforementioned Shapiro delay and light deflection is provided in Sec. \ref{sec:higher_order}.

\subsection{Retardation correction}\label{sec:z_ret}
The equations for $z_\mathrm{long}$ and $z_\mathrm{rel}$ (Eqs. (14) - (16)) assume that the reception event is known, in order to be able to evaluate all quantities.
However,
due to the finite propagation speed of light the reception occurs at a later, generally unknown,
time than the emission.
This means that the EOP needs to be solved to calculate the redshift based on the above equations.
In this section, we split the frequency shift into an instantaneous part, and a contribution due to the signal's retardation.

In flat spacetime
the propagation delay is simply given by the speed of light $c$ and distance $d = ||\bm{r_B} - \bm{r_A}||$ between emitting and receiving event, $\delta t = d / c$.
As discussed in \cite{linetTimeTransferFrequency2002} and Sec. \ref{sec:higher_order}, the general relativistic contributions to the propagation time yield contributions to the frequency shift of higher order, $\mathcal{O}(v^{3}/c^{3})$, which we neglect.
The same holds for the change in the light's tangent vector $\bm{\hat{k}}$ due to gravitational light deflection.

While the signal is propagating, the position and velocity of the receiver are changing
\begin{equation}
	\bm{\delta v_{B}} = \bm{a_{B,0}} \delta t + \mathcal{O}(\delta t^{2}),
	\quad  \bm{\delta r_{B}} = \bm{v_{B, 0}} \delta t + \mathcal{O}(\delta t^{2}),
\end{equation}
where $\bm{v_{B,0}}$ is the receiver's spatial velocity at time of emission and $\bm{a_{B,0}} = \dv{\bm{v_{B}}}{t}\big\vert_0 $ the acceleration 
Since the position of the receiver is changing during propagation, the unit ray vector connecting emitter and receiver naturally differs from that which would be obtained in the case of instantaneous transmission $\bm{{N}}_{0}
 =
\frac{\bm{r_{B,0}}-\bm{r_A}}{|\bm{r_{B,0}}-\bm{r_A}|}
$
\begin{equation}
	\begin{aligned}
		\bm{{N}}_{AB} &= \frac{\bm{{N}}_{0} + \bm{\delta r_{B}}/d} {\sqrt{1 + 2 \bm{{N}}_{0}\cdot \bm{\delta r_{B}}/d + \bm{\delta r_{B}}^{2}/d^{2}}} + \mathcal{O}\left(\delta t^2\right)
		      \\&= \bm{{N}}_{0} \left(1 - (\bm{{N}}_{0} \cdot \bm{v_{B, 0}}) \frac{\delta t}{d} \right) + \bm{v_{B, 0}} \frac{\delta t}{d}
	 + \mathcal{O}(\delta t^{2})
		\\&= \bm{{N}}_{0}\left( 1 - \bm{{N}}_{0} \cdot \frac{\bm{v_{B, 0}}}{c} \right) + \frac{\bm{v_{B, 0}}}{c} + \mathcal{O} \left(\frac{v^{2}}{c^{2}}\right)
	\end{aligned}
\end{equation}
We can then decompose the longitudinal Doppler shift into an instantaneous and a retardation contribution
\begin{equation}
	\label{eq:doppler_long_retardation}
	z_{\mathrm{long}} = z_{\mathrm{long}, 0} + z_{\mathrm{ret}},
\end{equation}
where
\begin{equation}
	\label{eq:z_long_0}
	z_{\mathrm{long}, 0} = \frac{1}{c} \left( (\bm{v_{B, 0}} - \bm{v_{A, 0}})\cdot \bm{{N}_{0}} \right)
\end{equation}
and
\begin{equation}
	\label{eq:z_ret}
	\begin{aligned}
	z_\mathrm{ret} =& \frac{1}{c^{2}} \big[
		\bm{v_{B,0}} \cdot (\bm{v_{B,0}} - \bm{v_{A,0}}) \\&
		- \bm{{N}_{0}} \cdot (\bm{v_{B,0}} - \bm{v_{A,0}})\,  \bm{{N}_{0}} \cdot \bm{v_{B,0}}
	\\
	&\quad 
	+ d (\bm{a_{B, 0}}
    \cdot \bm{{N}_{0}}) \big]
	+ \mathcal{O}\left( \frac{v^3}{c^{3}} \right),
	\end{aligned}
\end{equation}
where the first two lines in Eq. \eqref{eq:z_ret} correspond to the change in the signals tangent vector, and the last line to the change in the receiver's velocity vector.
The overall correction $z_{\mathrm{corr}}$ to the instantaneous longitudinal Doppler shift is then given by the (instantaneous) relativistic and retardation contributions
\begin{equation}
	\label{eq:z_corr}
	z_{\mathrm{corr}} = z_{\mathrm{ret}} + z_{\mathrm{rel, 0}},
\end{equation}
where $z_{\mathrm{rel, 0}}$ is the contribution from the instantaneous relativistic frequency shift
\begin{equation}
	\label{eq:z_rel_0}
	z_{\mathrm{rel}, 0} = \frac{1}{c^{2}} \bigg[
	\frac{1}{2}(v_{A,0}^{2} - v_{B,0}^{2}) 
	+ \left( \frac{GM}{r_{A,0}} -  \frac{GM}{r_{B, 0}} \right)
	\bigg].
\end{equation}

For an inter-satellite link, we use for the acceleration the Newtonian law of gravitation $\bm{a_{B,0}} = - \frac{GM}{r_{B,0}^3} \bm{r_{B,0}}$, while for an equatorial geostation the acceleration is given by Earth's angular velocity, $\bm{a_{B,0}} = -\omega_{E}^{2} \bm{r_{B,0}}$.

The above results show that, remarkably, the effect of retardation in the signal's propagation causes frequency shifts on the same order of magnitude as the lowest order relativistic effects, gravitational redshift and transversal Doppler shift.
This implies that any measurement or experiment that is accurate up to this order, must include relativistic as well as contributions from the light propagation.
Indeed, the contributions from retardation \eqref{eq:z_ret} include terms that are formally similar to the relativistic corrections, those include terms quadratic in the receivers velocity (\emph{transversal Doppler shift}) and terms proportional to the gravitational potential (\emph{gravitational redshift}).
We want to point out, that the frequency shift due to retardation, Eq. \eqref{eq:z_ret}, is not symmetric between up- and downlink configurations.
This is in contrast to the purely relativistic contributions, Eq. \eqref{eq:z_rel}.

It is interesting to investigate the behaviour of the gravitational terms of Eqs. \eqref{eq:z_ret} - \eqref{eq:z_corr} in the special cases of large and small separation, i.e. $d/r_{A,0} \gg 1$ resp. $d/r_{B,0} \ll 1$ and instantaneous radial transmission, i.e.
$\bm{{N}_0} \cdot \bm{r_{B,0}} = \pm r_{B,0}$,
for inter-satellite links:

i)
In the case $d/r_{A, 0} \gg 1$, i.e. $r_{B,0} \approx d$, of large distance between emitter and receiver, one finds
\[
GM \left(
\frac{1}{r_{A,0}} - \frac{1}{r_{B,0}} \mp \frac{1}{r_{B,0}}
\right),
\]
where the upper sign indicates uplink and the lower sign downlink.
Meaning that in the downlink scenario, the gravitational time dilation at the receiver, $-GM/r_{B,0}$, cancels exactly with the contribution from the retardation effect.

ii) In the case of small separation, $d/r_{A,0} \ll 1$, one finds by Taylor expansion of $\frac{1}{r_{A, 0}} = \frac{1}{r_{B,0} - d}$
\begin{equation}
\begin{split}
	GM \left( \frac{1}{r_{A,0}} -  \frac{1}{r_{B,0}} - \frac{d}{r_{B,0}^{3}} \bm{{N}_{0}}\cdot \bm{r_{B,0}} \right)
\\= 
GM\left(
\frac{d}{r_{B,0}^2} \mp \frac{d}{r_{B,0}^2}
\right)
+ \mathcal{O}\left(\left( \frac{d}{r_{B,0}}\right)^2\right)
,
\end{split}
\end{equation}
such that all gravitational terms cancel in the uplink scenario up to first order, while they double in the downlink case.

\subsection{Relativistic corrections to satellite orbits}
\label{sec:rel_orbits}
The previous analysis does not take relativistic corrections to the satellite orbits and their orbital velocities into account.
Assuming that the satellites are ideal test bodies free of any external forces, they move on geodesics of the underlying spacetime.
We assume the spacetime to be well approximated by the static, spherically symmetric Schwarzschild spacetime, which in Boyer--Lindquist coordinates takes the form
\begin{equation}
	\label{eq:metric_ss}
	\begin{aligned}
	g =& -\left( 1 - \frac{2GM}{c^{2}}\frac{1}{r} \right) dt^{2}
	+ \left( 1 - \frac{2GM}{c^2} \frac{1}{r} \right)^{-1} dr^{2}
	\\&+ r^{2} (d \theta^{2} + \sin^{2}(\theta) d \phi^{2}),
\end{aligned}
\end{equation}
where $M$ is the mass of Earth.
Using the Schwarzschild spacetime to model the gravitational field outside Earth, we neglect the higher order moment of Earth's gravitational potential such as the quadrupole moment, as well as perturbations due to the further bodies in the solar system.
Orbital perturbations are discussed at length in standard textbooks on celestial mechanics, e.g. in \cite{bate1971fundamentals, moulton2012introduction}.
While such perturbations must be included in a complete treatment, their omission here serves to isolate the relativistic effects we wish to investigate.
For a timelike geodesic in the Schwarzschild spacetime, the differential orbit equation as obtained from the geodesic equation \eqref{eq:geod} is \cite{straumannGeneralRelativity2013}
\begin{equation}
	\label{eq:eom_relativistic}
\frac{d^{2} u}{d \varphi^{2}} + u = c^2 \frac{r_\mathrm{S}}{2 L^{2}} +  \frac{3}{2} r_\mathrm{S} u^{2},\quad u = \frac{1}{r}, 
\end{equation}
where $r_\mathrm{S} = \frac{2 GM}{c^{2}}$ is the Schwarzschild radius.
The specific angular momentum $L$ and energy $E$ are constants of motion, where the angular momentum is related to the angular velocity by $L = r^{2} \dv{\varphi}{\tau}$.
The last term on the right hand side of Eq. \eqref{eq:eom_relativistic} is the relativistic contribution to the orbital dynamics.
The corresponding dimensionless quantity,
$\frac{3}{2} r_\mathrm{S} u = \frac{3r_\mathrm{S}}{2r}$,
for the Earth can be estimated to be
\begin{equation}
	\frac{3r_\mathrm{S}}{2r} \sim  \frac{\SI{0.01331}{m}}{r} 
    \begin{cases}
        \sim 2\cdot10^{-9}, \text{ for } r=r_{ISS}\\
        \sim 5\cdot10^{-10}, \text{ for } r=r_{GPS}
        \\
        \sim 3\cdot10^{-10}, \text{ for } r=r_{geo}.
    \end{cases}
\end{equation}
Since relativistic corrections to satellite orbits around Earth are small compared to the leading Newtonian dynamics, a perturbative approach is appropriate. 
Expanding to first order in the relativistic correction yields the modified orbital solution to first order in $r_\mathrm{S}$ \cite{straumannGeneralRelativity2013}
\begin{equation}
\begin{aligned}
	u(\varphi) = &\frac{1}{p} ( 1 + \varepsilon \cos \varphi) \\
			   & + \frac{3 r_{S}}{2 p^{2}} \left(1 + \frac{\varepsilon^{2}}{2} - \frac{\varepsilon^{2}}{6}\cos 2\varphi + \varepsilon \varphi \sin \varphi \right),
\end{aligned}
\end{equation}
where the last term is the relativistic correction.
The semi-latus rectum $p = a(1-\varepsilon^{2})$, semi-major axis $a$ and eccentricity $\varepsilon$ are those associated with the corresponding Keplerian ellipse.
The validity of the approximation requires that the semi-latus rectum is large compared to the Schwarzschild radius, i.e. $\frac{r_\mathrm{S}}{p}\ll 1$.
Within the relativistic correction, there is a constant radial offset, periodic terms and also a secular contribution arising from the last term.
The secular contribution leads to the well-known periapsis shift, which over one orbital period is given by
\cite{straumannGeneralRelativity2013, MTW_gravitation}
\begin{equation}
	\label{eq:peri_shift}
	\delta \varphi = \frac{3 \pi r_\mathrm{S}}{p }.
\end{equation}

The radius and radial velocity including first order relativistic correction are then
\begin{equation}
	\begin{aligned}
	r(\tau) &= 
	\frac{p}{\varepsilon \cos (\varphi )+1}
	\\&\quad - \frac{r_\mathrm{S}}{4} \frac{6 + 3 \varepsilon^2 - \varepsilon^2 \cos (2 \varphi )  + 6 \varepsilon \varphi  \sin (\varphi )}{(\varepsilon \cos (\varphi )+1)^2},
	\end{aligned}
\end{equation}
\begin{equation}
	\begin{aligned}
		\dot r (\tau) &= \dv{r}{u}\dv{u}{\varphi} \dv{\varphi}{\tau} = - L u^{\prime} (\varphi)
		\\
		&= L \bigg( \frac{\varepsilon}{p} \sin \varphi
		\\
		& \quad -\frac{3 r_{S} \varepsilon}{2p^{2}} \left( \frac{\varepsilon}{3} \sin 2\varphi + \sin \varphi  + \varphi \cos \varphi \right)
	  	\bigg).
	\end{aligned}
\end{equation}
For given angular momentum $L$, the angular velocity is $\dot{\phi} (\tau) = L / r^{2}$.
To express the velocities with respect to Schwarzschild coordinate we use \cite{straumannGeneralRelativity2013}
\begin{equation}
	\dot{t}(\tau) = \frac{E}{1 - \frac{r_{S}}{r}}
\end{equation}
and find
\begin{equation}
\begin{aligned}
	\frac{dr}{dt} &=  \frac{\dot{r} (\tau)}{\dot{t}(\tau)} = \dot{r}(\tau) \frac{1}{E} \left(1-\frac{r_{S}}{r} \right),
	\\
	\frac{d \phi}{dt} &= \frac{\dot{\phi}}{\dot{t}} =
	\frac{L}{E} \frac{1}{r^{2}} \left( 1 - \frac{r_{S}}{r} \right).
\end{aligned}
\end{equation}
Using the Newtonian relation between velocity and radius for Keplerian orbits
\begin{equation}
	\label{eq:vis_viva}
	v^{2} = GM \left( \frac{2}{r} - \frac{1}{a} \right),
\end{equation}
we find
that the relativistic corrections to the orbital velocity of the Kepler ellipse are of the order $\mathcal{O}(\frac{r_{S}}{r}) = \mathcal{O}\left( \frac{v^{2}}{c^{2}} \right)$.

Having found the first order relativistic corrections to position and velocity, we can estimate that the effect on the frequency shift, as per Eqs. \eqref{eq:z_total}-\eqref{eq:z_rel}, is at most of order
$\mathcal{O}\left( \frac{v^{3}}{c^{3}} \right)$.
In investigating the first order relativistic contributions to the frequency shift, we can therefore safely neglect relativistic corrections to orbital mechanics.

With Eq. \eqref{eq:vis_viva}, we can conveniently express the relativistic corrections to the relative frequency shift as a function of radius $r$ and semi-major axis $a$
\begin{equation}
	\label{eq:z_rel_kepler}
	z_\text{rel} = \frac{GM}{c^{2}} \left( \frac{2}{r_{A}} - \frac{2}{r_{B}} - \frac{1}{2 a_{A}} + \frac{1}{2 a_{B}} \right)
    + \mathcal{O}\left(\frac{v^3}{c^3}\right).
\end{equation}
Since the relativistic frequency shift \eqref{eq:z_rel_kepler} is only non-constant for elliptic orbits, it is also called `eccentricity effect' \cite{ashby2003relativity}.

If either receiver or emitter is a ground station fix on Earth's surface, Eq. \eqref{eq:vis_viva} obviously does not apply. Let e.g. $A$ be the ground station, then
\begin{equation}
	\label{eq:zrel_ground}
	z_\text{rel} = \frac{(\omega_\mathrm{E} R_\mathrm{E})^{2}}{2 c^{2}} + \frac{GM}{c^{2}} \left(\frac{1}{R_\mathrm{E}} + \frac{1}{2 a_B} - \frac{2}{r_B}\right),
\end{equation} where $\omega_\mathrm{E} = \frac{2 \pi}{\si{24}{h}}$ is the rotational frequency of Earth.
For circular orbits, the relativistic frequency shift relative to a ground station vanishes at a radius of
\begin{equation}
	\label{eq:zero_rel_shift}
	R_{*} = \frac{3}{2} \left( \frac{(\omega_\mathrm{E} R_\mathrm{E})^2}{2 GM} + \frac{1}{R_\mathrm{E}} \right)^{-1}
	\approx \SI{9551}{km}
	\approx \frac{3}{2} R_{\mathrm{E}}.
\end{equation}

\section{Examples}
\label{sec:examples}
Here we calculate the frequency shift and secret key capacities for explicit configurations of ground stations and satellites.
We consider ground station to satellite, as well as inter-satellite up- and downlinks.
To illustrate the significance of the effects presented in Sec. \ref{sec:freq_shift}, we investigated representative orbits of various heights and eccentricities but restricted to the equatorial plane.
Those include low/medium Earth orbits (LEO/MEO) and geosynchronous and geostationary orbits (GSO/GEO)
with eccentricities up to $\varepsilon = 0.7$ (highly elliptical orbits, HEO, e.g. Molniya orbit).
While only examples, they cover many possible configurations and illustrate the significance in similar experimental setups.

The start of the signal exchange is chosen to occur when both communicating parties are at their respective perigee and at the same azimuth (i.e. are initially aligned).
The plots have been obtained by simulating satellites to be on Keplerian orbits of given semi-major axis $a$ and eccentricity $\varepsilon$ or equatorial ground station and the light rays to propagate in a flat spacetime.
The frequency shifts were calculated according to Eqs. \eqref{eq:z_ret}-\eqref{eq:z_rel_0}.
From this, the PLOB bound was calculated using Eq. \eqref{eq:cap_plob}.

We assume the signal to have a peak frequency to bandwidth ratio of $R = 10^{10}$ which commonly occurs in quantum communication applications
\cite{Soh_selfreferenced_cvqkd_llo_2015,Bing_Locallocaloscillator_CVQKD_2015,Huang_locallocaloscillator_cvqkd_2015,Liao2025,marieSelfcoherentPhaseReference2017a}.
As discussed in Fig. \ref{fig:PLOB}, it is at this magnitude where relativistic effects on the frequency shift start to become relevant.

To isolate the relativistic and retardation effects on the key rates, we assume that the longitudinal Doppler shift, as given by the instantaneous velocities and positions at signal emission, is corrected.
Therefore, the mode mismatch is only caused by relativistic contributions and retardation.

\subsection{Satellite to ground station}
\subsubsection{Geosynchronous satellite}

\begin{figure*}[t]
\centering
\begin{minipage}[b]{0.49\textwidth}
    \centering
	\includegraphics[width=1\linewidth]{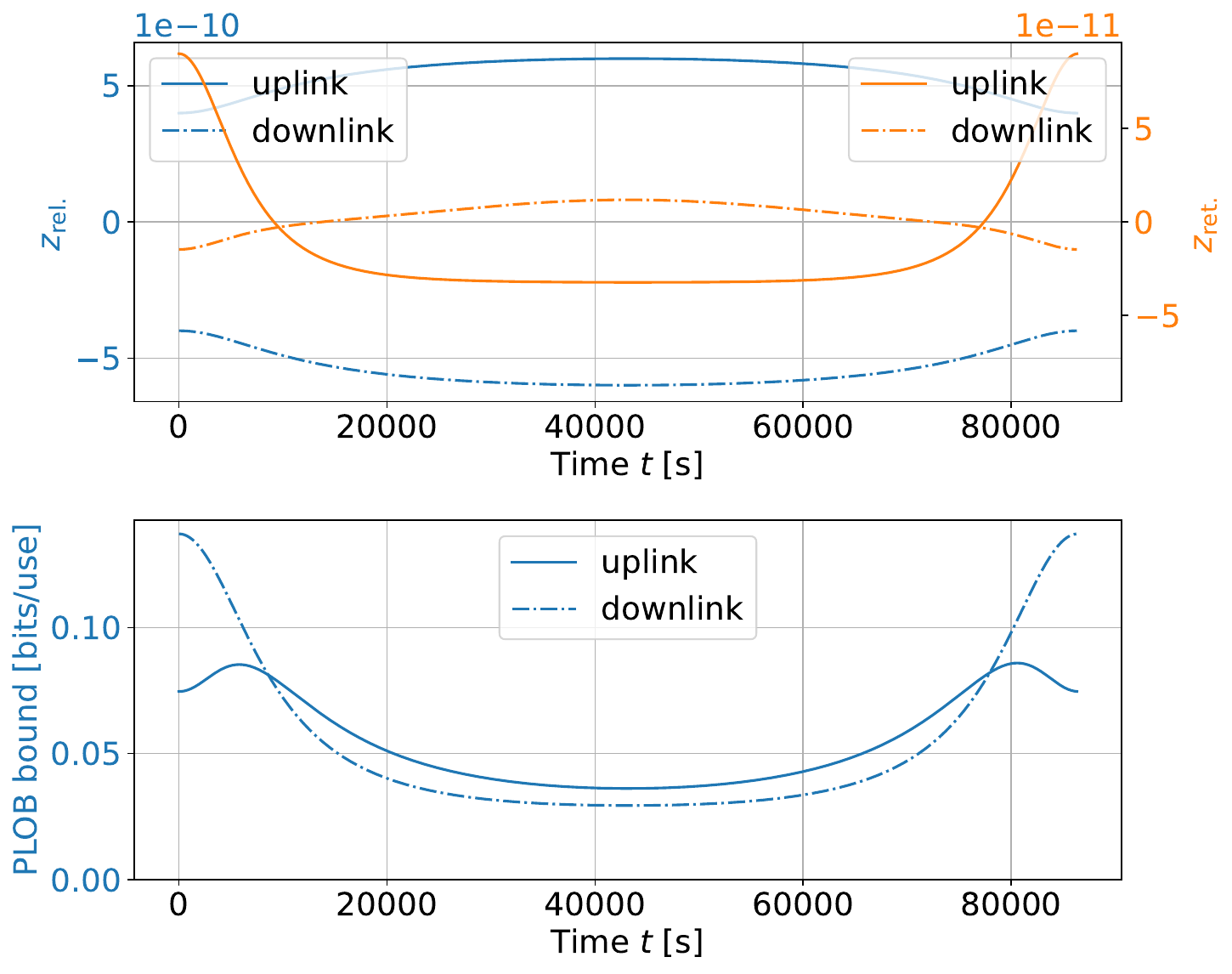}
\end{minipage}
\hfill
\begin{minipage}[b]{0.49\textwidth}
    \centering
	\includegraphics[width=1\linewidth]{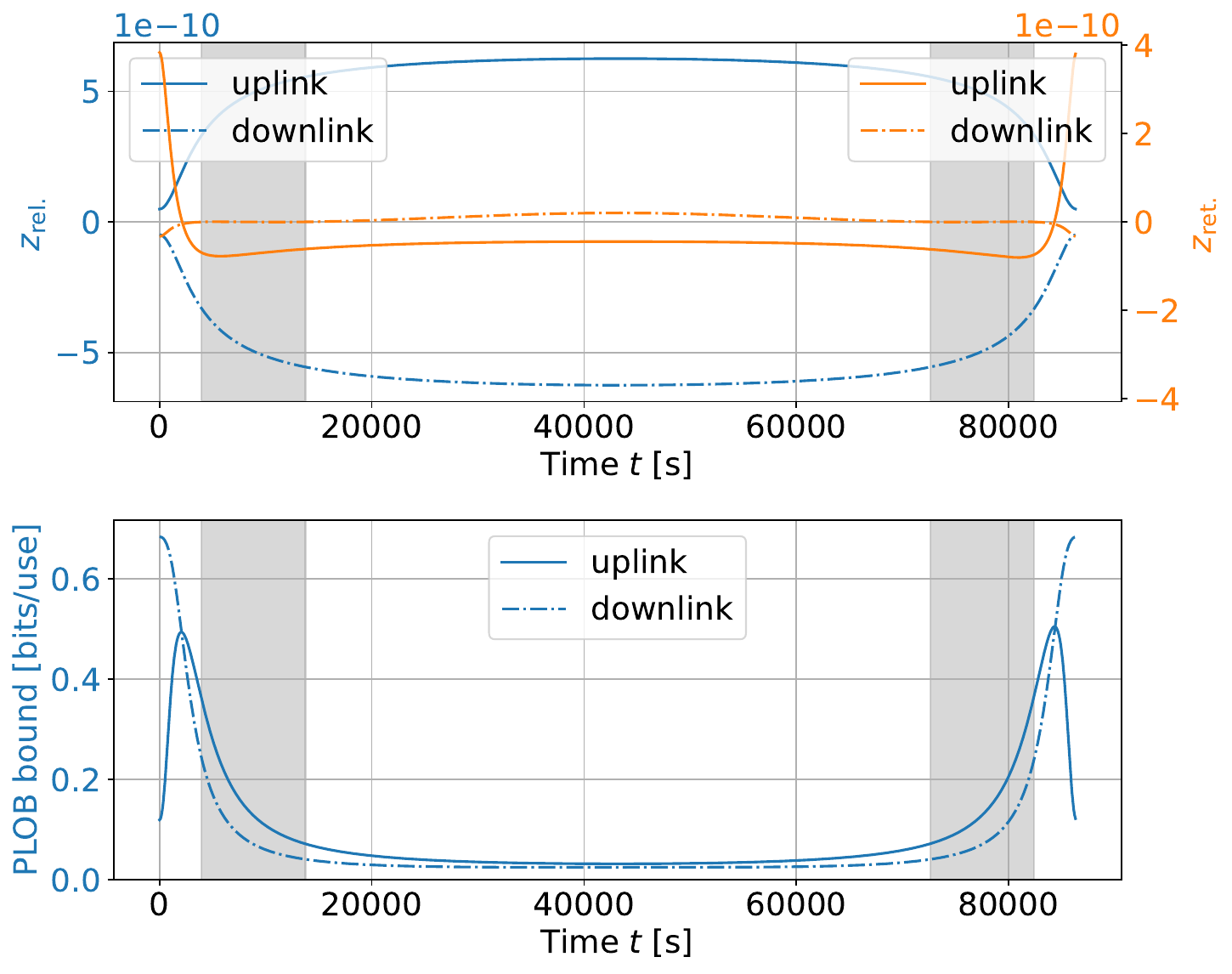}
\end{minipage}
\caption{
Top: Evolution of the relativistic frequency shift (blue) and retardation effect (orange) for an equatorial ground station and an equatorial geosynchronous orbit with eccentricity $\varepsilon = 0.4$ (left) and $\varepsilon = 0.7$ (right) over one day for up- and downlink. 
Gray areas indicate times without line of sight. %
In the uplink, the frequency shift due to the retardation effect (Sec. \ref{sec:z_ret}) is about an order of magnitude smaller than the relativistic frequency shift for the lower eccentricity orbit.
For the higher eccentricity orbit, the retardation effect is comparable in magnitude to the relativistic contributions and exceeds them near perigee.
For the downlink, the retardation effect is generally less significant due to the lower velocity of the ground station compared to the satellite. The retardation effect is generally extremal at perigee and apogee due to, respectively, large velocities and distances.
Bottom: The corresponding PLOB bound for a lossy channel with loss given by the above frequency shifts as calculated from Eqs. \eqref{eq:wamb_gauss}--\eqref{eq:cap_plob} and baseline transmissivity of $\eta_{0} = 0.4$.
Due to the retardation, the PLOB bound has a local minimum at perigee
(e.g. at $t=0$ and $t = \SI{24}{h} = \SI{86 400}{s}$)
in the uplink which is not present in the downlink due to the less significant retardation effect.
Generally, the PLOB bound peaks at (uplink) or around (downlink) perigee and drops of toward apogee, where it reaches a global minimum. For more eccentric orbits the peaks and drops in the PLOB bound are more significant, leading to more effective but shorter windows of large capacity.
}
\label{fig:ground_geosynch}
\end{figure*}

First, we consider geosynchronous satellites above an equatorial ground station.
In the special case of a geostationary satellite,
the ground station naturally has constant line of sight and the relative frequency shift is constant at $z = z_{\mathrm{rel}} \approx \SI{\pm 5.398e-10}{}$ for up- and downlink respectively, which is purely the relativistic contribution from gravitational and relativistic Doppler shift. 
The longitudinal Doppler shift $z_\mathrm{long}$ is exactly zero, as well as the retardation effect $z_\mathrm{ret}$.
This is the case for any configuration of ground station and geostationary satellite because the satellite appears stationary from the point of view of the ground station.

Orbits of non-zero eccentricity remain geosynchronous (i.e. orbital period of 24 hours) but become elliptical and are no longer seen as stationary from Earth.
The altitude at perigee decreases as the eccentricity increases.
The frequency shift now varies as the velocity as well as the radius change as per Eq. \eqref{eq:zrel_ground}.
For eccentricities of $\varepsilon = 0.4$ and $\varepsilon = 0.7$ the orbits as well as frequency shifts are shown in Fig. \ref{fig:ground_geosynch} over a period of $\SI{24}{h}$.
The relativistic contributions to the frequency shift 
are extremal at the apogee where the gravitational redshift is largest.
At the perigee, due to the increasing satellite velocity, the transverse Doppler shift in part compensates the gravitational redshift leading to a minimal frequency shift (resp. maximal for downlink).
In the uplink, the retardation effect is maximal at the perigee, $z_{\mathrm{ret}} \approx \SI{8.98e-11}{}$, 
and plateaus quickly toward the apogee at approximately $z_{\mathrm{ret}} \approx \SI{-3.23e-11}{}$ with a zero crossing in between.
In the downlink the retardation effect is overall less significant, which is intuitively clear since the, now emitting, geosynchronous satellite moves at a larger total velocity than the receiving ground station.
For the same reason the retardation effect does not exhibit the strong peak at the perigee. However, the frequency shift is slightly larger in magnitude at perigee since the emitters velocity enters through the projection on the modified wavevector $k$ as per Eq. \eqref{eq:z_ret}.

These frequency shifts have significant influence on the achievable communication performance via the PLOB bound which is directly calculated from the total mode mismatch via Eqs. \eqref{eq:wamb_gauss} and \eqref{eq:cap_plob}.
Since in this example the retardation effect is most of the time significantly smaller than the relativistic frequency shift,
the PLOB bound mirrors mainly the relativistic frequency shifts behaviour.
Except for at the perigee, where, in the uplink, the PLOB bound has a local minimum at perigee due to the retardation effect adding onto the relativistic contribution
, which is not present in the downlink scenario.
The features just discussed become more pronounced for larger eccentricities, as seen in Fig. \ref{fig:ground_geosynch} (right).
The PLOB bound achieves larger peaks but also drops off much steeper, allowing for shorter but more effective windows of communication.
The magnitude of drop in capacity as well as the duration of this window depends strongly on the eccentricity of the satellite's orbit, as seen by the much steeper drop in Fig. \ref{fig:ground_geosynch}.
Additionally, the line of sight between satellite and ground station occasionally breaks for high eccentricity geosynchronous orbits, since the slow satellite velocity at the apogee leads to the ground station moving out of the satellite's view due to Earth rotation.
Towards the perigee the satellite catches up with the ground station because of its increasing velocity.

\subsubsection{LEO and MEO satellites}

\begin{figure*}[t]
\centering
\begin{minipage}[b]{0.49\textwidth}
    \centering
	\includegraphics[width=1\linewidth]{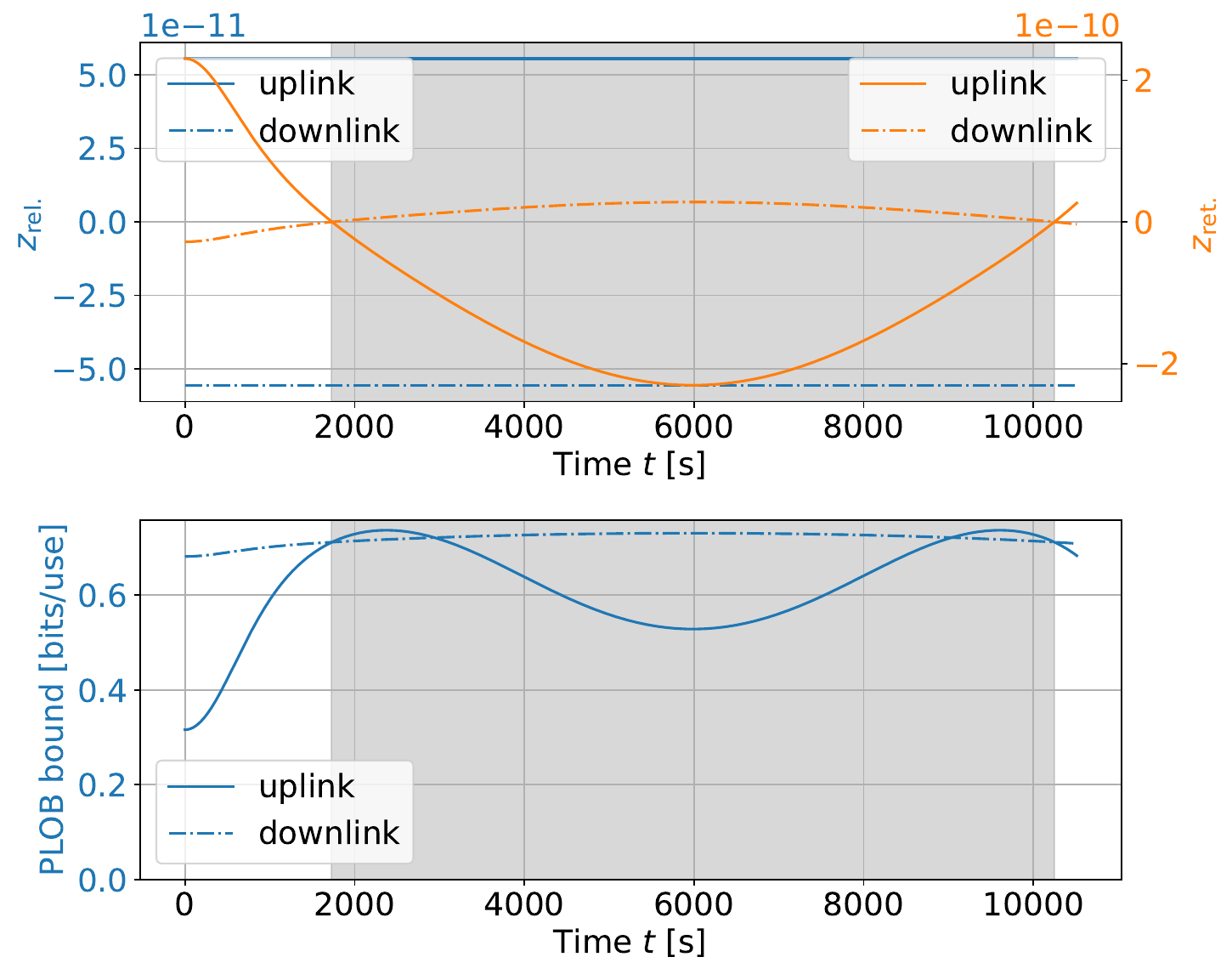}
\end{minipage}
\hfill
\begin{minipage}[b]{0.49\textwidth}
    \centering
	\includegraphics[width=1\linewidth]{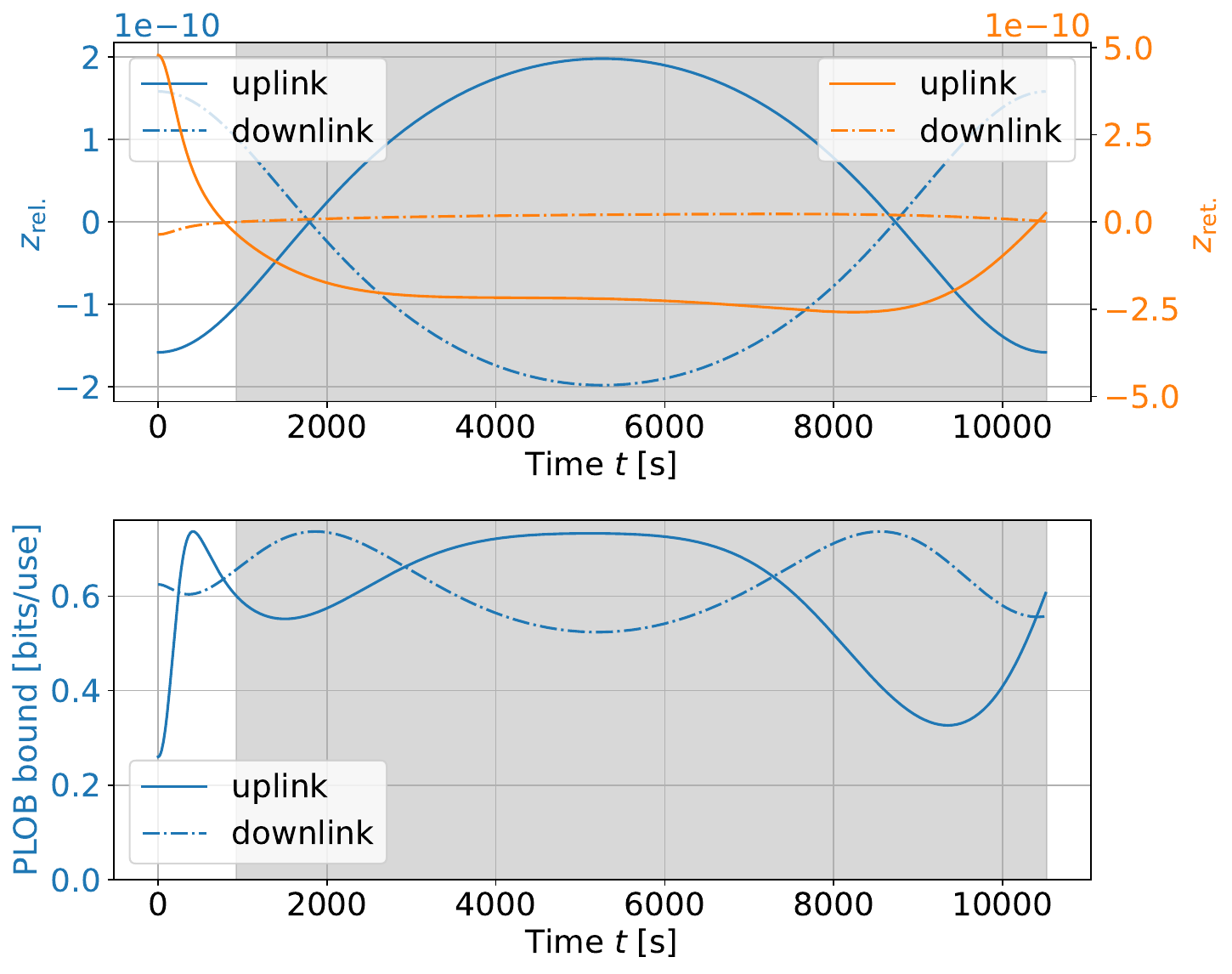}
\end{minipage}
\caption{
Top: Evolution of the relativistic frequency shift (blue) and retardation effect (orange) for an equatorial ground station and an equatorial MEO satellite 
with a semi-major axis of $a = \SI{10 378}{km}$
and eccentricities $\varepsilon = 0.0$ (left) and $\varepsilon = 0.2$ (right) over one orbital period of $\SI{10511}{s}$.
Left: The line of sight is interrupted (gray) after about $\SI{\sim 1726}{s}$ and picks up again toward the end of the satellite's orbital period.
Since the orbit is circular and therefore radii and speeds of ground station and satellite are constant, the relativistic contributions to the frequency shift are constant as well.
The retardation effect is, as in Fig. \ref{fig:ground_geosynch}, more significant in the uplink and almost an order of magnitude larger than the downlink.
Since the relativistic frequency shift is constant, the PLOB bound's dynamic is given purely by the retardation effect. It remains nearly constant for the downlink, while in the uplink the PLOB bound is minimal at perigee and increases toward the end of line of sight.
Right:
The non-zero eccentricity leads to non-constant and increased relativistic contributions, which are extremal at perigee and apogee where the transversal and gravitational frequency shift dominate, respectively.
In the uplink, the retardation effect drops off faster and plateaus toward the apogee.
The duration of the line of sight is reduced to about $\SI{936}{s}$.
The PLOB bound at perigee is reduced compared to the circular orbit due to the increased retardation effect. In the uplink, the PLOB bound quickly peaks within line of sight and drops off towards the end of line of sight.
}
\label{fig:ground_MEO}
\end{figure*}

Satellites on low and medium Earth orbit (LEO and MEO) have larger velocities than higher orbital satellites and are therefore naturally expected to suffer more significant Doppler shifts and retardation effects.
In Fig. \ref{fig:ground_MEO}, the frequency shifts and PLOB bound are shown for links between ground stations and a MEO satellite with a semi-major axis of $a=\SI{10 378}{km}$ and eccentricities of $\varepsilon = 0.0$ and $\varepsilon = 0.2$ (heights at perigee $h_p = \SI{4000}{km}$ and $h_p = \SI{1924.4}{km}$).
Results for different altitudes (e.g. lower LEO or higher MEO) do not differ qualitatively.
A further effect of practical significance is the increasingly short duration of line of sight for orbits of low altitude as e.g. seen in the examples of Figs. \ref{fig:ground_MEO}, where line of sight is lost already after $\SI{\sim1000}{s}$.
For zero eccentricity, the relativistic frequency shift remains constant but its magnitude is decreased.
For an altitude of \SI{4000}{km} the frequency shift is about an order of magnitude smaller than the geostationary satellite.
Similarly to the geosynchronous orbits of Figs. \ref{fig:ground_geosynch}, the uplink retardation effect is again much more significant.

Another feature distinguishing LEO from MEO orbits appears in combination with eccentricity, where the relativistic frequency shifts now changes sign, as in Fig. \ref{fig:ground_MEO} (right).
As the satellite approaches its perigee, its velocity increases while the radius decreases such that
at some point the transverse Doppler shift exactly compensates the gravitational redshift.

\subsection{Satellite to Satellite}

Having investigated ground station to satellite communication, we now turn to inter-satellite links.
The major difference being that a ground station is moving on a non-geodesic circular path while the satellites are assumed to be freely falling.

\begin{figure}
	\centering
	\includegraphics[width=1\linewidth]{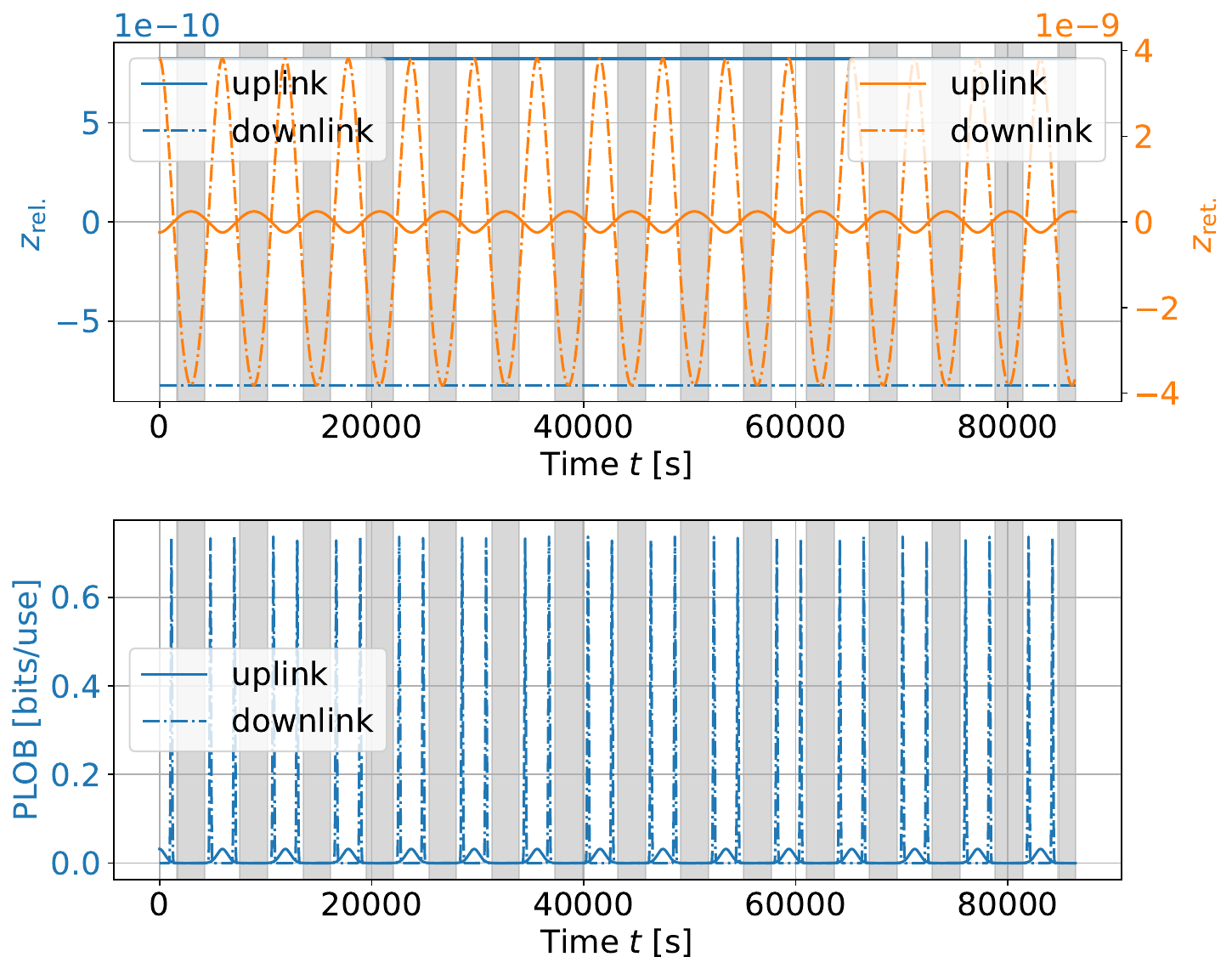}
	\caption{
	Top: Evolution of the relativistic frequency shift (blue) and retardation effect (orange) for an equatorial inter-satellite link between a circular LEO satellite (altitude $h_\mathrm{leo}=\SI{400}{km}$, eccentricity $\varepsilon_\mathrm{leo}=0$) and a geostationary orbit ($h_\mathrm{geo} \approx \SI{35870}{km}$, $\varepsilon_\mathrm{geo}=0$) over one day for up- and downlink.
	Gray regions indicate regions where there is no line of sight between the satellites.
	The retardation effect oscillates at the orbital frequency of the LEO satellite while the relativistic frequency shift remains constant since both satellites are on zero eccentricity orbits.
	In contrast to the ground station--satellite links of Figs. \ref{fig:ground_geosynch}, \ref{fig:ground_MEO}, the retardation effect is much more significant in the downlink due to the larger orbital velocity of the LEO satellite. In the downlink, the retardation effect is an order of magnitude larger than the relativistic frequency shift.
	Bottom: The corresponding PLOB bound for a lossy channel with loss given by the above frequency shifts as calculated from Eqs. \eqref{eq:wamb_gauss}--\eqref{eq:cap_plob} and baseline transmissivity of $\eta_{0} = 0.4$.
	The PLOB bound inherits the oscillatory behaviour of the retardation effect. In the uplink, the PLOB bound peaks at perigee while in the downlink it peaks very sharply before and after perigee (again, inverse to the results of Figs. \ref{fig:ground_geosynch}, \ref{fig:ground_MEO}).
	}
	\label{fig:sat_sat1}
\end{figure}

The exchange between a ground station and a geostationary satellite (GEO) therefore differs significantly from a LEO to GEO link,
which is depicted for a circular LEO at an altitude of $h = \SI{400}{km}$ in Fig. \ref{fig:sat_sat1}.
Due to the short orbital period combined with the low altitude of LEO satellites (about \SI{1.5}{h} at altitude \SI{400}{km}), the line of sight is very limited. 
While the relativistic contribution $z_\text{rel}$ stays constant because the satellites' radii and speeds are constant, the frequency shift due to retardation, $z_\mathrm{ret}$, is oscillating around zero. Here, the downlink oscillates with much larger amplitude than the uplink and is inverted in sign. This is in contrast to the ground station-satellite links, where the uplink suffered more significant retardation effect (for geosynchronous and lower orbits).
This is reflected in the PLOB bound which inherits the oscillatory behaviour of the frequency shift.
As expected from the frequency shift, the peaks in the uplink are located where the retardation effect is minimal and therefore counteracts the relativistic frequency shift.
Since the downlink is an order of magnitude larger than the relativistic frequency shift, however, the PLOB bound peaks very narrowly shortly before and after perigee and quickly plateaus toward apogee. In the downlink there are therefore two maxima in the key rate per orbital period.

\begin{figure*}[t]
\centering
\begin{minipage}[b]{0.49\textwidth}
    \centering
	\includegraphics[width=\linewidth]{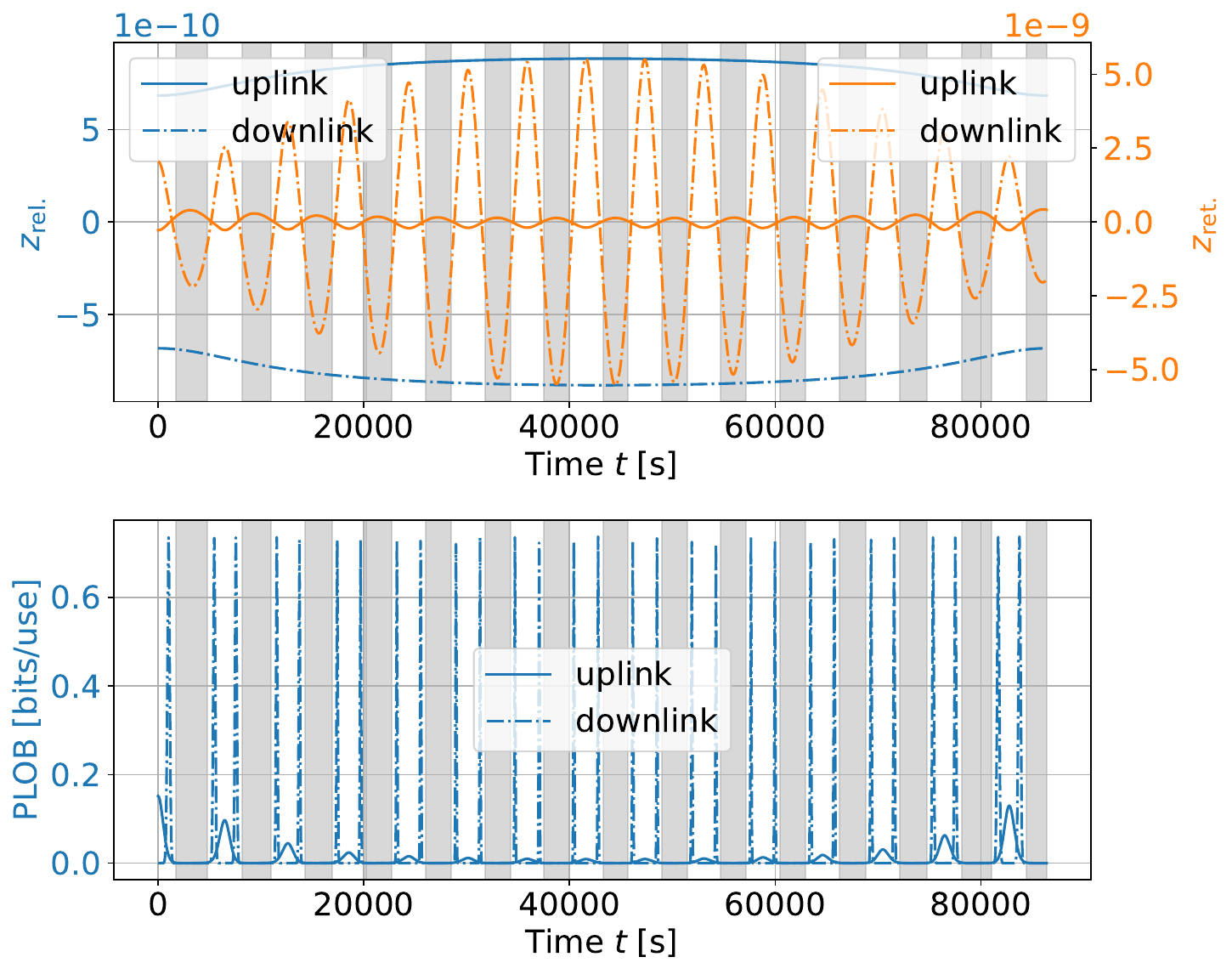}
\end{minipage}
\hfill
\begin{minipage}[b]{0.49\textwidth}
    \centering
    	\includegraphics[width=\linewidth]{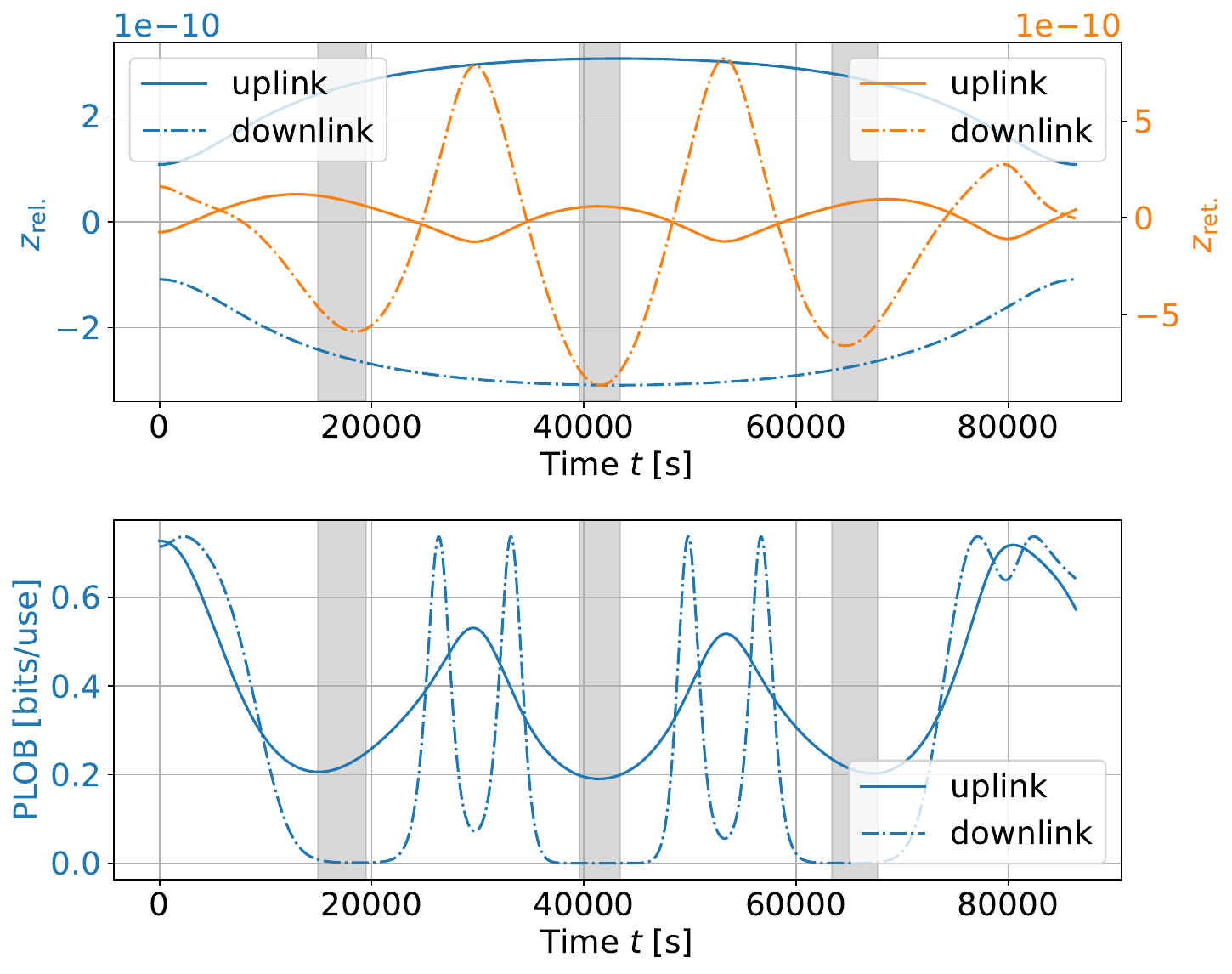}
\end{minipage}
\caption{
	Top: Evolution of the relativistic frequency shift (blue) and retardation effect (orange) for an equatorial inter-satellite link between a circular LEO satellite (altitude $h_\mathrm{leo}=\SI{400}{km}$, eccentricity $\varepsilon_\mathrm{leo}=0$) (left),  and an elliptical geosynchronous orbit ($a_\mathrm{geo} \approx \SI{42248}{km}$, $\varepsilon_\mathrm{geo}=0.4$) over one day for up- and downlink.
	Right: The inner LEO orbit is replaced by a MEO orbit (altitude $h_\mathrm{meo}=\SI{10000}{km}$, eccentricity $\varepsilon_\mathrm{meo}=0$).
	For the LEO orbit (left), the frequency shift and PLOB bound behave similarly to Fig. \ref{fig:sat_sat1} except for the now dynamic relativistic frequency shift caused by the eccentricity of the outer orbit. The amplitude of the retardation effect is now also changing along the orbit: In the downlink it is maximal at apogee, in the uplink at perigee.
	For the MEO orbit (right), the oscillations naturally occur on a larger time scale due to the increased orbital period and the retardation effect decreases significantly.
	Bottom:
	The corresponding PLOB bound for a lossy channel with loss given by the above frequency shifts as calculated from Eqs. \eqref{eq:wamb_gauss}--\eqref{eq:cap_plob} and baseline transmissivity of $\eta_{0} = 0.4$.
	The PLOB bound reflects the oscillatory behaviour of the retardation effect, the amplitude of oscillation is given by the relativistic contribution which, for uplink, is smallest at perigee and largest at apogee, where largest key rates are possible. For the downlink, however, the amplitude is not affected by the change in relativistic frequency shift since it reaches its maximum value (as given by baseline transmissivity $\eta_{0}=0.4$) due to the sufficiently compensating contribution from the retardation effect.
}
\label{fig:sat_sat2}
\end{figure*}

In Fig. \ref{fig:sat_sat2} we consider inter-satellite links between a satellite on a circular inner orbit (left: LEO at $h_\mathrm{LEO} = \SI{400}{km}$, right: MEO at $h_\mathrm{MEO} = \SI{10 000}{km}$) and an outer satellite on an elliptical geosynchronous orbit (eccentricity $\varepsilon = 0.4$ ).
As before, when eccentricity is introduced, the relativistic contribution to the frequency shift becomes dynamic, as depicted in Fig. \ref{fig:sat_sat2}.
The eccentricity leads to an additional periodic variation in the amplitude (oscillating envelope) of the retardation effect $z_\mathrm{ret}$ on the time scale of the period of the outer eccentric orbit. For the uplink the amplitude is maximal at the perigee, for the downlink at the apogee.
For the downlink, the retardation effect cancels the relativistic frequency shift exactly around the perigee.

For an inner orbit at a higher altitude, as depicted in Fig. \ref{fig:sat_sat2} (right), the line of sight windows are naturally significantly longer and oscillations occur at lower frequency due to a larger orbital period of the inner satellite.
Since the satellites tend to be closer, the relativistic frequency shift is smaller, although it still leads to a significant drop in the PLOB bound at large separation.
In the downlink, the frequency shift from the retardation effect is also decreased due to lower orbital velocity of the inner satellite.
Overall, the MEO--GEO link allows for larger PLOB bounds (in the uplink) and longer windows of communication than the LEO--GEO link.

\section{Higher order corrections}
\label{sec:higher_order}

Finally, it is worth considering the limitations of the approximations applied here to the emitter--observer problem.
To this end, the example satellite setups presented in the previous section are treated in the general relativistic framework, with satellite and signal trajectories in the Schwarzschild spacetime as solutions to the geodesic equation, and the corresponding redshift is compared with the above results.
For a more thorough explanation of the procedure, see Appendix \ref{APP:rel_corr}.

\begin{figure}
	\centering
	\includegraphics[width=1\linewidth]{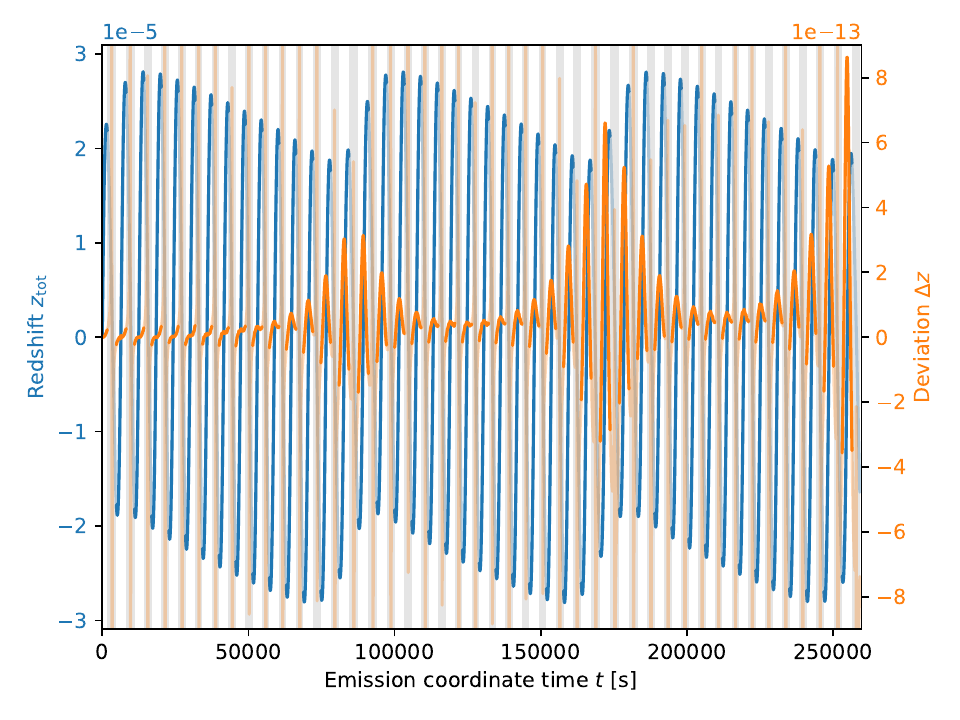}
	\caption{Total redshift (blue) calculated in GR and redshift deviation (orange) caused by approximating satellite and signal trajectories in the EOP; the satellite initial conditions correspond to Fig. \ref{fig:sat_sat2} (left). Regions in which line of sight is lost are shaded with a gray background. On timescales of the LEO orbital period, light bending dominates the deviation, while on timescales of the geosynchronous orbital period, periapsis shift causes a redshift deviation whose maximal amplitude increases over each full geosynchronous satellite revolution.}
	\label{fig:higher_order_corr}
\end{figure}

How the redshift deviates between the two frameworks is exemplified in Fig. \ref{fig:higher_order_corr} for an uplink inter-satellite signal exchange between a satellite on a circular equatorial low Earth orbit and an eccentric geosynchronous orbit (compare Fig. \ref{fig:sat_sat2}, left).
The blue curve describes the redshift calculated numerically without approximations, while the orange curve quantifies the difference in redshift (deviation) arising from applying the approximations to the satellite and signal trajectories.
In the numerical calculations, the overall oscillating redshift behavior also present in Fig. \ref{fig:sat_sat2} (left) is reproduced.
Simultaneously, a deviation in the redshift starts to develop, oscillating with a time-variable amplitude that alternates between monotonously in- and decreasing.
The deviation curve's (orange) maxima outside of the shaded regions can be enveloped by a function whose local maxima increase over time.
While the frequency of the deviation curve's oscillations coincides with the circular LEO period, the envelope's maxima coincide with the geosynchronous orbital period.
The deviation's behavior is caused by the two main effects that are neglected by the approximations described in Sec. \ref{sec:freq_shift}; light bending and periapsis shift.
Early in the simulation, when the periapsis shift is still small, light bending effects dominate at this order of magnitude even far away from the Earth center.
The oscillations in the deviation curve are caused by the LEO satellite passing the point of radial transmission, where the signals are bent in one direction before passing it and in the opposite direction after passing it.
At the point of radial transmission the light bending (as well as the longitudinal Doppler shift) becomes negligible, wherefore the deviation strongly decreases.
However, after multiple LEO periods, a maximum starts to emerge at the point of radial transmission.
The behavior of the envelope tracing these maxima is ascribed to the cumulative effect of the periapsis shift and the position of the geosynchronous satellite along the orbit.
The envelope's local maxima coincide with the moment when the geosynchronous satellite reaches the orbit's periapsis, where the velocity along the respective orbit is largest.
Since the periapsis shift induces a rotation of the tangent vector to the orbit when compared to the Kepler ellipse, and since the vectors' magnitudes are largest at the periapsis, so is the deviation in the redshift.
Once the satellites leave the periapsis, the deviation decreases accordingly until the apoapsis is reached.

Therefore peaks in the deviation are to be expected for satellite setups with large satellite velocities, e.g. from highly eccentric orbits.
If the signal propagation time is exceedingly large, the cumulative periapsis shift together with a high eccentricity may lead to deviations in the redshift that are of the order of relativistic effects discussed in this work.
In this case the approximations are no longer sensible, see for example in Fig. \ref{fig:z_diff} in Appendix \ref{APP:rel_corr} for a comparison of satellite setups presented in the previous section.
Rather, a numerical approach using the complete GR framework as described in Appendix \ref{APP:EOPEA} is then necessary.
Altogether a discussion of relativistic effects on QKD in the manner shown here requires a careful choice of satellite setups.

\section{Discussion and conclusion}
\label{sec:conclusion}

We investigated the relativistic and retardation effects to the relative frequency shift in the context of space-based optical quantum communication.
Regarding the relativistic effects, we demonstrated that to order $\mathcal{O}(v^2/c^2)$ only the well-known gravitational redshift and the transversal Doppler shift contribute, that can be attributed to general and special relativistic time dilation.
Using the recently introduced relativistic numerical software package \textit{GREOPy} \cite{hackstein2025, hackstein2024} for solving the emitter--observer problem for arbitrary orbital constellations, we investigated higher order corrections to the frequency shift from relativistic orbital dynamics, such as perigee shifts, as well as light bending effects on the signal transmission. We found that on the timescale of a few orbital periods, these effects can be neglected for relevant communication setups. For combinations of very large semi-major axis and eccentricity however, higher order effects may become relevant and can, for example, be investigated with GREOPy \cite{hackstein2025}. 

Additionally, we investigated the impact of the finite propagation time of light on the frequency shift.
We decomposed the longitudinal Doppler effect into an instantaneous contribution and a retardation contribution, where the latter is of the same order of magnitude as the relativistic effects, i.e. $\mathcal{O}(v^2/c^2)$.
Therefore, it must be taken into account as well whenever relativistic effects are relevant.
While the leading order relativistic contributions depend only on radius and speed, the retardation effect is more complex in that it depends on the velocity's and light's directions.
It is asymmetric in up- and downlink configurations and in inter-satellite links generally more significant in the downlink, since the inner satellite's velocity is greater (for circular orbits). While the retardation effect is by definition included in the usual expressions for the longitudinal Doppler effect, eq.~\eqref{eq:z_long}, we think it is both convenient and instructive to work with the instantaneous quantities used in this paper.

To analyze the effect on communication, we have modeled the frequency shift-induced mode mismatch in a CV-QDK protocol as a lossy quantum channel, whose capacity is determined by the PLOB bound \cite{pirandola2017fundamental}.
We assumed that the leading-order longitudinal Doppler shift had been properly accounted for and corrected.
Whether relativistic effects affect communication depends largely on the peak frequency-to-bandwidth ratio of the signal carrier.
The order of magnitude of the combined relativistic and retardation effects on the frequency shift in the Earth's gravitational field is about $z \sim 10^{-10}$, such that they are significant for peak frequency-to-bandwidth ratios of $R \gtrsim 10^{10}$.
Using numerical simulations, we have confirmed this for explicit examples of equatorial constellations of ground station-satellite and inter-satellite links.

This investigation offers several ways for further research.
An extension of the equatorial two-party links to arbitrary orbits and ground stations and even to more complex constellations, and eventually to quantum networks, is a natural next step.
In such networks, entanglement will be an important resource for communication and computation, as such, relativistic frequency shifts might also play an important role in space-based distribution of entanglement.
In this regard the investigation of the significance of frequency shifts on different communication protocols, especially ones using entanglement as a resource \cite{Ekert1991QuantumCB, yin2017satellite, Yin2020, Wang_2021, Lu_cao_micius_space_review_2022} might be worthwile.
While atmospheric effects are significant in such networks, as shown in e.g. \cite{pirandolaLimitsSecurityFreespace2021, meister2025simulation}, the integration of relativistic and atmospheric effects into a unified framework might be of relevance to high-precision sensing applications.

Relativistic effects beyond frequency shifts might soon be important to experimental studies using, for example, deep space quantum links \cite{Mohageg2022_deep_space_quantum_link}.
Those include the deformation of spatial modes, tidal effects, and gravitational phases accumulated along propagation, as studied in e.g. \cite{exirifard_distortion_photon_2022, toward_communication_curved_exirifard_2021}.
An operational investigation of this requires a framework which models the three-dimensional propagation of the signal in curved spacetime as well as the emission and reception, which will be published in a future article.

\section*{Acknowledgements}

We thank Dennis Philipp, Marian Cepok, Altin Shala, David Edward Bruschi, and Andreas Wolfgang Schell for useful comments and discussions.
E.S. and D.R. acknowledge funding by the Federal Ministry of Education and Research of Germany in the project “Open6GHub” (grant number: 16KISK016).
We further acknowledge funding by the Deutsche Forschungsgemeinschaft (DFG, German Research Foundation) under Germany’s Excellence Strategy – EXC-2123 QuantumFrontiers – 390837967 as well as within the CRC TerraQ – Project-ID 434617780 – SFB 1464 and the Research Unit RU 5456 Clock Metrology - Project-ID 490990195.

\bibliographystyle{apsrev4-2}
\bibliography{ref}

\clearpage
\newpage

\appendix

\section{\label{APP:EOPEA} Emitter--observer problem}

\begin{figure}[t]\centering
  \includegraphics[width=0.9\linewidth]{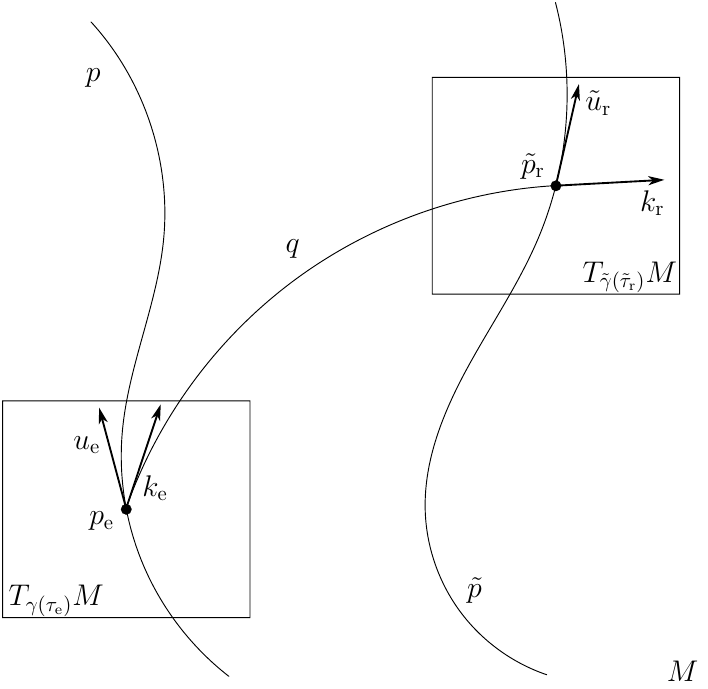}
  \caption{Two satellites on timelike curves traced by \(p, \tilde{p} \in M\) communicate by transmitting a light signal at the emission event marked by \(p_\text{e}\), which travels along a lightlike geodesic traced by \(q\in M\) and is received at the reception event marked by \(\tilde{p}_\text{r}\). The emitter--observer problem is finding \(\zeta(s)\) and the reception event for a given emission event. The signal's frequency as measured by emitter and observer can then be calculated from the tangent vectors \(u_\text{e} = \dot{\gamma}(\tau_\text{e}), \tilde{u}_\text{r} = \dot{\tilde{\gamma}}(\tilde{\tau}_\text{r}), k_\text{e,r} \coloneq \dot{\zeta}(s_\text{e,r})\) at the events, respectively.}
\label{fig:eop_sketch}
\end{figure}

The emitter--observer problem (EOP) can be expressed geometrically in the following way:
Assume two satellites move along timelike curves \(\gamma(\tau)\), \(\tilde{\gamma}(\tilde{\tau})\), respectively, parametrised with proper time \(\tau, \tilde{\tau}\).
Further assume that a light signal transmitted between the satellites travels along a lightlike curve \(\zeta(s)\) parametrized with an affine parameter \(s\).
The image sets of the curves \(p\coloneq\text{Img}(\gamma), \tilde{p}\coloneq\text{Img}(\tilde{\gamma}), q\coloneq\text{Img}(\zeta)\) are shown in Fig. \ref{fig:eop_sketch}.
For a light signal exchange, define an emission event \(\gamma(\tau_\text{e})\) with \(p_\text{e}\coloneq\text{Img}(\gamma(\tau_\text{e}))\), as well as a reception event \(\tilde{\gamma}(\tilde{\tau}_\text{r})\) with \(p_\text{r}\coloneq\text{Img}(\tilde{\gamma}(\tilde{\tau}_\text{r}))\) that are lightlike separated.
Then fixing the emission event automatically fixes the reception event and the lightlike curve \(\zeta(s)\) connecting them.
A continuous light signal exchange between the satellites in the interval \((\tau_1, \tau_2)\) is only possible if each emission event is lightlike separated with a reception event.
The EOP now consists of finding the unique \(\zeta(s)\) such that the equations
\begin{align}
    \gamma(\tau_\text{e}) = \zeta(s_\text{e}), && \tilde{\gamma}(\tilde{\tau}_\text{r}) = \zeta(s_\text{r}).
    \label{EQ:EOP_cond}
\end{align}
are fulfilled.

For the purposes of this work, the EOP is solved numerically within the GR framework with the help of the \textit{GREOPy} software package written in the Python language.
It can integrate the geodesic equation for lightlike geodesics in arbitrary spacetimes and given arbitrary timelike worldlines that allow for Eqs. \eqref{EQ:EOP_cond} to be fulfilled, meaning there exist emission and reception events that are pairwise lightlike separated.
The general idea of the numerical procedure including how \textit{GREOPy} solves the EOP will be outlined in this section.

First the satellite setup must be chosen, e.g. a ground station and a LEO satellite or two satellites on different orbits.
One fixes the orbits by specifying the corresponding initial conditions \(x^\mu(\tau), u^\mu\coloneq\dot{x}^\mu(\tau)\) (event and four-velocity components) and the propagation times;
in this paper, discussion is restricted to the equatorial plane, so initial conditions contain \(\theta = \frac{\pi}{2}, \dot{\theta} = 0\), and the propagation time is restricted to a couple of orbital periods.
Solving the geodesic equation numerically for these initial conditions, in this case using SciPy's \cite{2020SciPy-NMeth} solve\_ivp function, one obtains one timelike geodesic acting as the emitter and one as the observer.
Each point \(\gamma(\tau_\text{i})\) on the emitter's timelike curve can act as an emission event for the light ray, meaning one immediately demands \(\gamma(\tau_\text{e}) = \zeta(s_\text{e})\).
As stated above, this already fixes the solution of the EOP, and now it needs to be calculated.

One way to find the lightlike geodesic fulfilling the second condition \(\tilde{\gamma}(\tilde{\tau}_\text{r}) = \zeta(s_\text{r})\) can be found via the shooting method \cite{keller2018numerical}, on which the \textit{GREOPy} algorithm is based.
It expresses a boundary-value problem in a number of initial-value problems, of which one also fulfills the boundary conditions.
Since the emission event is fixed, the initial four-velocity \(k_\text{e}\coloneq\dot{x}(s_\text{e})\) is varied until a resulting lightlike geodesic crosses the receiver's wordline; in other words, where the receiver's wordline intersects the emission event's light cone.

For an operational approach, it is convenient to define a local tetrad \((e_\mu)\) at the emission event in order to express \(k_\text{e}\) in terms of celestial angles \((\Phi, \Psi)\) on the emitter's celestial sphere.
Instead of varying the initial four-velocity, one can then scan the emitter's celestial sphere for the pair of celestial angles which yield a lightlike geodesic that satisfies \eqref{EQ:EOP_cond}.
This is done by taking advantage of the differential evolution algorithm, specifically the SciPy implementation based on \cite{storn1997differential}.
It is a global optimization method able to handle possibly non-differentiable functions of multiple variables.
It is applied to the EOP by defining a function that 1) calculates a lightlike geodesic for a chosen pair of celestial coordinates and 2) calculates the closest approach of the corresponding solution to the receiver worldline in a hypersurface \(t=\text{const.}\) with coordinate time \(t\).
In this case, the \textit{closeness} of two events is defined as the spacelike distance between them in the hypersurface;
for weakly-curved regions of spacetime, the spacelike distance defined by the metric may be replaced by the euclidean definition of distance in flat spacetime to shorten calculation time.
Since the reception event most likely does not coincide with one of the reception orbit sampling points, the distance is calculated for each light ray sampling point with a linearly interpolated reception orbit until the closest approach is found.
In weakly-curved regions of spacetime, the distance function is monotonous with only one minimum.
Over multiple iterations, a number of lightlike geodesics are calculated over the initial conditions interval, and the closest approaching lightlike geodesic is kept as a reference for the next iteration to restrain the initial conditions (IC) interval.
A new population of lightlike geodesics is then calculated in the slightly smaller IC interval, then compared with each other, with the IC of the closest light ray iteratively serving to reduce the IC interval until a certain threshold for the spacelike distance of closest approach is crossed.
The final lightlike geodesic serves as the solution to the EOP by satisfying \eqref{EQ:EOP_cond} up to the given threshold.
After the light ray curve \(\zeta(s)\) and its four-velocity \(k(s)\) along the curve are completely determined, the redshift between emitter and observer can be calculated.
Repeating this for each step along the emission orbit shows the redshift behavior over time as seen e.g. in Figs. \ref{fig:ground_geosynch} - \ref{fig:ground_MEO}.

\section{\label{APP:rel_corr} Numerical comparison of (non-) relativistic orbital mechanics and signal propagation}

\begin{figure*}[t]
    \centering
	\includegraphics[width=.9\textwidth]{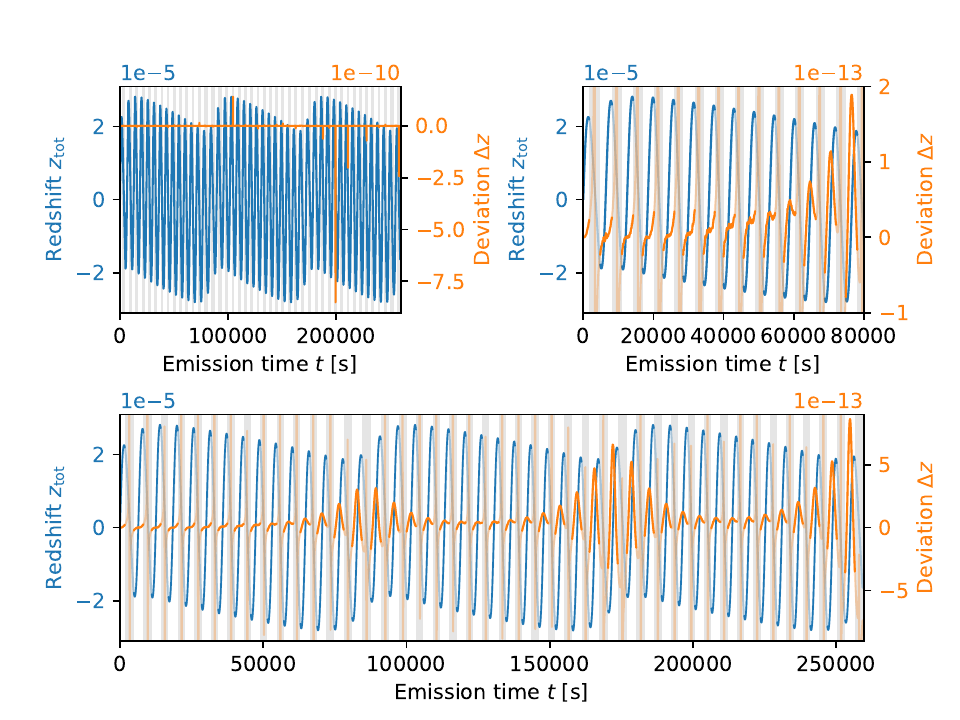}
    \caption{Total redshift (blue) calculated in GR and redshift deviation (orange) caused by approximating satellite and signal trajectories in the EOP; the satellite initial conditions correspond to Fig. \ref{fig:sat_sat2} (left). Top left: The presence of light deflection strongly increases the redshift deviation in non line of-sight regions \(r \ll r_\text{Earth}\) (gray background), where stronger light deflection is present. Bottom: Scaling the axis w.r.t. deviation values within line of sight regions, the deviation's oscillating behavior is visualized. Top right: Restricting the emission time \(t\) to one geosynchronous orbital period shows more clearly the emergence of a maximum in the deviation caused by the periapsis shift.}
    \label{fig:los_discussion}
\end{figure*}

The discussion of the main text focuses on the influence of relativistic corrections to the redshift up to orders including \(\mathcal{O}\left( \frac{v^2}{c^2} \right)\), neglecting higher order relativistic corrections to Kepler orbits, which are estimated to be at most of the order \(\mathcal{O}\left( \frac{v^{3}}{c^{3}} \right)\), and approximating the signal transmission with straight lines on a flat background metric.
This is in contrast to a general relativistic treatment of the EOP, where receiver and light signal move on a timelike curve and lightlike geodesic, respectively.
The purpose of this appendix is to elaborate on and quantify the deviation in the redshift simulation between the general relativistic treatment and the approximations adopted in the main text, and by extension, to justify the adopted approximations regarding the scope of this work.
For the discussion, the EOP is solved and the redshift over time calculated, once in the general relativistic framework and once with the above-mentioned approximations, for the satellite configurations presented in Figs. \ref{fig:ground_geosynch} - \ref{fig:ground_MEO} and \ref{fig:sat_sat1} - \ref{fig:sat_sat2}.

First concerning the satellite motion.
A Kepler orbit is generally described by six orbital elements, but since discussion is restricted to movement contained within the equatorial plane (\(\theta = \frac{\pi}{2}\)) of the spherically-symmetric spacetime, this number is reduced to three;
the orbital shape is completely determined by the orbital period \(T\) and the eccentricity \(\varepsilon\), while the curve's starting event is fixed by the mean anomaly at epoch \(M_0\).
A satellite moving along this orbit travels on a non-geodesic curve \(\gamma(t)\) that can be expressed in Schwarzschild coordinates, where the radius and true anomaly (azimuthal angle) are given by
\begin{align}
    \begin{split}
	   r(t) &= a \cdot \left(1 - \varepsilon \cdot \text{cos}(E(t))\right), \\
        \nu(t) &= 2 \cdot \text{arctan}\left(\sqrt{\frac{1 + \varepsilon}{1 - \varepsilon}} \text{tan}\left(\frac{E(t)}{2}\right)\right)
    \end{split}
	\label{eq:Kepler_coordinates}
\end{align}
with the eccentric anomaly \(E(t)\) parametrized by coordinate time \(t\).
The spatial velocity vector \(v\) transported along the satellite curve on a Keplerian orbit is then given by the rate of change of radius and true anomaly.
In Newtonian gravity, the presence of absolute time fixes the parametrization of \(\gamma \) and \(v\) w.r.t. this absolute time.
Once this curve described by Eq. \eqref{eq:Kepler_coordinates} in Schwarzschild coordinates is embedded in the Schwarzschild spacetime, the parametrization of \(\gamma\) and \(v\) is determined up to the choice of time parameter with which to measure the spatial rate of change;
either w.r.t. coordinate time \(t\) or proper time \(\tau\).
The corresponding four-velocity components \(u^\mu\) along \(\gamma\) are given by \(u^\mu = \dv{x^\mu}{\tau} = \dv{x^\mu}{t}\dv{t}{\tau}\).
Now the choice lies in whether to measure the three-velocity along the curve as
\begin{align}
v^i = \dv{x^i}{t} \quad \text{or} \quad v^i = \dv{x^i}{\tau}.
\end{align}
The latter case is used e.g. for treatment that includes relativistic corrections to Newtonian equations of motion \cite{petit2010iers}, resulting in a satellite (e.g. on a circular orbit) feeling relativistic perturbation effects during propagation.
Here, the three-velocity is chosen as measured w.r.t. coordinate time.
Then the spatial velocity vector \(v(t)\) together with the normalization condition
\begin{align}
    -c^2 = g_{\mu \nu} \dot{x}^\mu \dot{x}^\nu
\end{align}
completely determine the satellite four-velocity \(u(\tau)\).

Instead of modeling the satellite's curve after a Keplerian orbit in Schwarzschild coordinates, one can simulate it as a geodesic on curved spacetime.
The curve's initial conditions can be calculated analogously as for the Kepler orbit with Eqs. \eqref{eq:Kepler_coordinates} and their derivatives.
Since here timelike curves start at the periapsis, as determined by \(M_0\), the initial radial velocity is always zero.
Solving the geodesic equation for these initial conditions yields a satellite curve in Schwarzschild coordinates which deviates from the Keplerian orbit of Newtonian gravity.

Second concerning the signal exchange.
As argued in Sec. \ref{sec:z_rel_contr}, the Earth's weak gravitational field may be characterized by a flat background metric \(\eta\) with some small perturbations.
By neglecting these perturbations, a signal can be calculated as propagating in the flat background spacetime and following a curve described by Schwarzschild coordinates.
The corresponding tangent covector \(k\) to the curve is transported using \(\eta\), and the associated frequency measured by the observer is calculated by projecting it onto the observer's four-velocity \(u\), see Eq. \eqref{eq:freq_shift2}.
On the other hand, by not linearizing gravity, the signal can be calculated as propagating along a geodesic in curved spacetime.
Then the corresponding tangent covector is transported using the general metric \(g\), and the associated frequency measured by the observer is calculated analogously.
Depending on the chosen method, the reception event will be different, contributing to the difference in redshift.

Solving the EOP once in the GR framework and once with the approximations discussed above, one can compare the redshift measurements to quantify the total deviation.

The first result is shown in Fig. \ref{fig:los_discussion} for an emitter on a circular equatorial LEO orbit and a receiver on an eccentric geosynchronous orbit, plotted over three periods of the reception orbit (see Fig. \ref{fig:sat_sat2} left for more details).
In each sublot, the blue curve describes the total redshift over time for the EOP solved numerically but without approximations, while the orange curve describes the deviation in redshift between the EOP solved with and without approximations.
Starting at the top left subplot, the redshift oscillation over one LEO orbit period (present in Fig. \ref{fig:sat_sat2}) is reproduced, and the oscillation behavior over multiple orbital periods of the reception orbit, hinted at in the main text figure, is visible more clearly.
The redshift deviation is dominated by occasional peaks with a varying magnitude;
the maximum magnitude of \(\sim 8\cdot10^{-10}\), which is of the order of relativistic effects discussed in the main text, appears to suggest that relativistic effects on the signal curve and receiver satellite are non-negligible.
However, the highlighted line of sight (los) between the satellites reveals these peaks to by lying in regions where the los is lost (gray background).
Indeed, the peaks in the deviation are caused by strong light bending, occurring when signals pass close to the Earth's center.
The bending angle in the Schwarzschild spacetime may be expressed by Einstein's deflection formula
\begin{align}
    \delta = \frac{4GM}{c^2r_\text{m}},
\end{align}
where \(r_\text{m}\) is the signal's minimal radius coordinate.
Thus, in regions with \(r_\text{m} \ll r_\text{Earth}\), the signal curve will deviate from the straight line approximation more strongly, resulting in a larger redshift deviation;
while the bending is estimated to be negligible in LEO orbit regions (\(\Delta z \approx 10^{-15}\) for \(r \approx \SI{7000}{km}\)), in regions close to the Earth center (e.g. \(r \approx \SI{700}{m}\)), the influence is estimated to be \(\frac{\SI{7000}{km}}{\SI{700}{m}} = 10^4\) orders of magnitude larger, which is reflected in the figure.
Since the signal path depends on the emission event, the numerical sampling of the emission curve also influences how close the signal passes to the Earth's center.
Together with the fact that this effect is not cumulative over multiple orbital periods, this leads to large differences between maxima in different non-los regions.
This work focuses on satellite communication, wherefore the rest of the discussion will be limited to regions with line of sight, which is shown in the right and bottom subplot of Fig. \ref{fig:los_discussion}.
The bottom subfigure shows the overall behavior for three periods of the reception orbit, while the top right subfigure shows the first period in more detail.
Here, since signals travel at \(r>r_\text{Earth}\), the light bending effects are much weaker.
Now the redshift deviation's oscillating behavior, regularly interrupted by non-los sections, is more visible.

The deviation, described by an oscillation with varying amplitude, whose extrema can be traced by an envelope, can be explained by two main causes.
Focusing first on emission times during the geostationary satellite's first orbital period (\(t < \SI{86400}{s}\)), see Fig. \ref{fig:los_discussion} top right, the first five line of sight sections (\(t < \SI{30000}{s}\)) are dominated by light bending since the periapsis shift is still small.
The moment of radial transmission between the two satellites, where the light bending is weakest and the longitudinal Doppler shift vanishes, constitutes the center in the line of sight section.
To the right of this center, when the satellites start to move in their respective directions along their orbits, the relative distance between the satellites grows and the direction of signal transmission starts to monotonically rotate away from the radial direction.
Simultaneously, the influence of light bending on the signal monotonically increases, which is shown by the increase in the deviation.
Analogously, to the left of the line of sight center, the satellites enter the line of sight section with a highly non-radial signal transmission, where the light bending is large, and with a decreasing relative satellite distance, resulting in a blueshift with relatively large magnitude.
Over time the satellites then approach the moment of radial transmission, while the signal transmission rotates more and more in radial direction;
during this time the effect of light bending monotonically decreases until it vanishes at the point of radial transmission.
After some LEO orbital periods, two extremal points emerge along the deviation curve around the point of radial transmission, becoming more and more pronounced over time.
This is explained by the periapsis shift that monotonically increases over time.
The tangent vector transported along a Keplerian orbit always points in the same direction when comparing the satellite after one and multiple orbital periods. 
If the orbital periapsis shifts over time instead, leading to orbital precession, a comparison between two satellite positions one or multiple orbital periods apart yields a difference in direction of the tangent vector, meaning the tangent vector transported along the geodesic orbit undergoes additional rotation.
When comparing a geodesic orbit with a Keplerian curve, this relative rotation between the velocity vectors induces a difference in the redshift measurements that monotonically increases with the periapsis shift and can best be seen at the point of radial transmission, where the Doppler shift vanishes.
At the periapsis, the geostationary satellite velocity, i.e. the tangent vector's magnitude, is largest along its orbit, and therefore the difference between the two velocity vectors in largest as well along one orbital period.
This can be observed well in the figure when focusing on the oscillation's enveloping function (highlighting the oscillation's extrema), where after one geostationary orbital period the deviation's magnitude starts to decrease again, see Fig. \ref{fig:los_discussion} bottom subfigure, because the geostationary satellite moves away from the periapsis, as its velocity diminishes.
This results in the envelope having local minima at the apoapsis and local maxima at periapsis.
Since the periapsis shift is a cumulative effect, the overall deviation becomes larger as time goes on, as can be seen in the envelope's three monotonically increasing maxima.
After a sufficiently long satellite travel time, the magnitude of the deviation will reach the order of magnitude of relativistic effects discussed in the main text, meaning the applied approximations break down.
This is however only if satellite trajectories are never corrected during flight, and the timescales considered here are orders of magnitude longer than timescales considered for actual line of sight measurements, meaning the approximations hold for this orbit constellation.

\begin{figure*}[t]
  \centering
  \includegraphics[width=.9\textwidth]{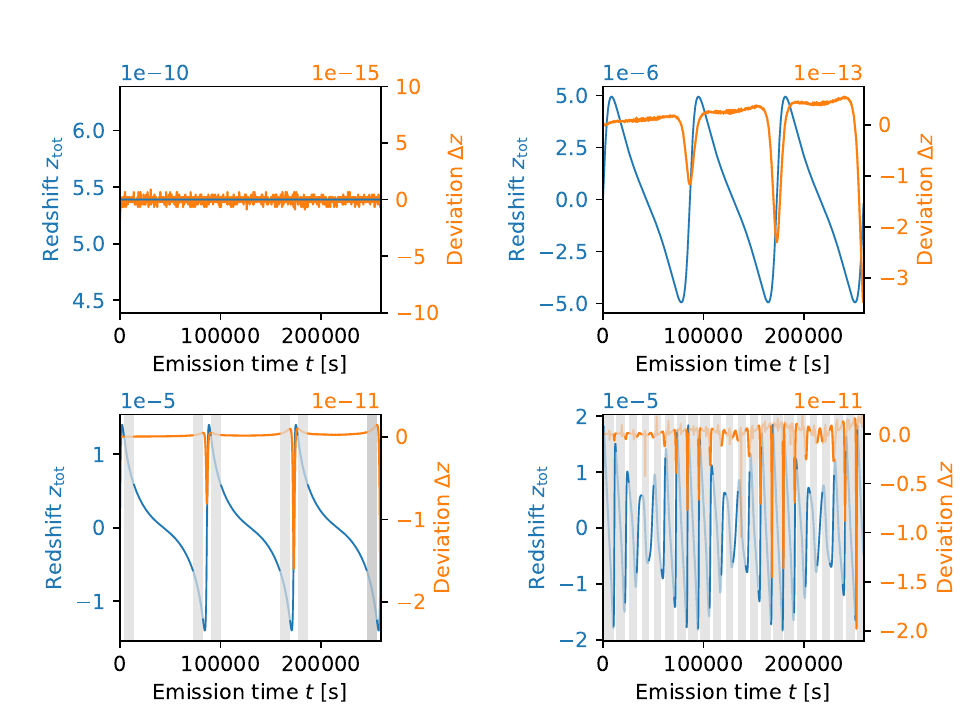}
  \caption{Total redshift (blue) calculated in GR and redshift deviation (orange) caused by approximating satellite and signal trajectories in the EOP. Gray background indicates times during which line of sight is lost. Top left: Signal exchange between a geosynchronous satellite above an equatorial ground station. Neither periapsis shift nor light bending influence the redshift, so the small deviation is due to numerical error. Top right: Satellite setup corresponds to Fig. \ref{fig:ground_geosynch} (left). The deviation increasing over time at the periapsis due to the periapsis shift can be seen. Since the deviation's overall magnitude is relatively small, numerical error around the apoapsis is resolved. Bottom left: Satellite setup corresponds to Fig. \ref{fig:ground_geosynch} (right). The only difference to the deviation in the top right subfigure is its increased magnitude due to increased eccentricity and therefore increased satellite velocity at the periapsis. Here, the deviation at the periapsis is of the magnitude of relativistic effects discussed in this work. Bottom right: Satellite setup corresponds to Fig. \ref{fig:ground_MEO} (right). The deviation's quick oscillating behavior on the timescale of the MEO period is due to light bending around the Earth, while the overall increase in the magnitude over multiple periods of the ground station is due to periapsis shift. Here again, the deviation at the periapsis is of the magnitude of relativistic effects.}
  \label{fig:z_diff}   
\end{figure*}

Fig. \ref{fig:z_diff} shows the redshift and redshift deviation for other orbit constellations discussed in the main text, analogously to Fig. \ref{fig:los_discussion} with light bending filtered out in non-los regions.
Except for the results shown in the first subfigure, very similar behavior can be seen, attributed to light bending and the tangent vector rotation due to periapsis shift.
In subfigure 1 the signal transmission is always radial for the satellite in geosynchronous orbit, leading to a small redshift deviation since light bending and periapsis shift are absent.
Its behavior can be traced back to numerical noise due to the EOP algorithm used in this work.
Beyond that, the magnitude of the redshift deviation is below the order of the relativistic effects discussed in the main text for only two of the shown subfigures.
In Fig. \ref{fig:z_diff} (bottom left), where signals are transmitted between a ground station and a highly eccentric geosynchronous orbit, the deviation peaks at \(\sim 10^{-11}\) after one orbital period.
Similarly, in Fig. \ref{fig:z_diff} (bottom right), where signals are transmitted between a ground station and an eccentric satellite orbit of \(\SI{4000}{km}\) height, the deviation peaks in the same order of magnitude after one orbital period of the ground station.
The reason has been discussed in the main text as well, where apart from a cumulative effect from long measurement times, the relative satellite velocity (caused by low altitudes and/or high eccentricity) plays an important role when choosing a satellite setup for quantum key distribution.
From this it is clear that careful consideration needs to be given to the satellite constellations when adopting these approximations for QKD.

\end{document}